\documentclass[11pt]{article}
\usepackage{amsmath,amsthm,amssymb}
\usepackage{enumerate}
\usepackage{hyperref}

\newtheorem{theorem}{Theorem}[section]
\newtheorem{lemma}{Lemma}[section]

\newtheorem{definition}{Definition}[section]
\newtheorem{remark}{Remark}[section]

\newcommand{\eoe}{\hfill$\lozenge$}

\newcommand{\dfn}{\stackrel{\Delta}{=}}

\newcommand{\calG}{{\cal G}}
\newcommand{\calM}{{\cal M}}
\newcommand{\calN}{{\cal N}}
\newcommand{\calT}{{\cal T}}

\newcommand{\calX}{{\cal X}}
\newcommand{\calY}{{\cal Y}}
\newcommand{\calV}{{\cal V}}
\newcommand{\calZ}{{\cal Z}}
\newcommand{\calA}{{\cal A}}

\newcommand{\E}{\mathbb{E}}

\newcommand{\LZ}{{\scriptstyle\mathrm{LZ}}}

\newcommand{\eps}{\epsilon}
\newcommand{\hHm}[1]{\hat{H}_{#1}}

\allowdisplaybreaks

\begin{document}

\title{Fano's Inequality and Channel Capacity in the\\ Individual-Sequence Regime}
\author{Neri Merhav\\[4pt]
\normalsize The Viterbi Faculty of Electrical and Computer Engineering\\
\normalsize Technion -- Israel Institute of Technology\\
\normalsize Haifa 3200003, Israel\\
\normalsize {\tt merhav@technion.ac.il}}
\date{}
\maketitle
\thispagestyle{empty}

\begin{abstract}
We derive an individual-sequence analogue of Fano's
inequality, replacing entropy by Lempel--Ziv (LZ)
complexity and probability of error by empirical error
frequency.
In the setting considered, an arbitrary encoder maps a message
to a channel input sequence $\{x_t\}$, the channel maps the input to an
output $\{y_t\}$ in a completely arbitrary manner, and a
block finite-state decoder produces an
estimate of the channel input from the output.
The individual-sequence Fano inequality lower-bounds the
empirical symbol-error frequency in terms of the LZ
complexity of the channel input conditioned on the channel output.
This leads to a channel coding converse: reliable
communication at a given rate is possible only if that
rate does not exceed the individual-sequence LZ channel
capacity, defined as the LZ mutual information maximized over
channel inputs and minimized (worst-case) over a noise set.
We also prove a matching achievability theorem for the class
of \emph{symbol-wise invertible} channels, in which
the channel output at each time $t$ is $y_t = f(x_t, v_t)$,
$v_t$ being a `noise' variable,
and $f(\cdot,v)$ is a
bijection for every fixed $v$.
This class includes group-additive channels
$y_t = x_t \oplus v_t$ studied by Feder and coworkers.
For such a channel, any rate below the LZ capacity
is achievable via a randomized code (with shared randomness
between encoder and decoder) with vanishing \emph{expected}
per-message error frequency,
simultaneously for every $v^n$ in a given noise set
$\mathcal{N}$, provided $\mathcal{N}$ is closed under
block permutations; this condition is satisfied
by any noise set that depends on $v^n$ only via its
type class with respect to such blocks. The achievability uses a universal decoder that requires no
knowledge of the channel and no feedback.\\

\end{abstract}

\section{Introduction}
\label{sec:intro}

\subsection{Background and Motivation}

Fano's inequality~\cite{Fano61} is one of the cornerstones of
classical information theory.
It states that for any communication system in which a
message $X$ is transmitted over a channel with output $Y$
and decoded as $\hat{X}$, the error probability $P_e \dfn \Pr\{\hat{X} \neq X\}$ satisfies
\begin{equation}
\label{eq:classical-fano}
    \phi(P_e)\dfn h_2(P_e) + P_e\log(|\calX|-1) \geq H(X|Y),
\end{equation}
where $H(X|Y)$ is the conditional entropy of $X$
given $Y$, $h_2(\cdot)$ is the binary entropy
function, and $|\calX|$ is the cardinality of the alphabet, $\calX$, of the
random variable $X$. 
This is the most common tool for proving converses to channel coding theorems.

In the \emph{individual-sequence} (or \emph{universal})
framework of Ziv and Lempel~\cite{ZL77,Ziv78,ZL78},
sequences are individual deterministic objects and no
probabilistic model is assumed on either the source or
the channel.
The role of conditional entropy rate is played by the
\emph{conditional LZ complexity}  --- the normalized
cost of compressing $x^n=(x_1,\ldots,x_n)$ given
$y^n=(y_1,\ldots,y_n)$ using the conditional version of the LZ78
algorithm~\cite{ZL78,
Ziv85,me00,UK03}.

No individual-sequence analogue of Fano's inequality has
appeared in the prior literature; this paper fills that gap
and then derives a complete channel coding theorem --- converse and
achievability --- in the individual-sequence framework.

\subsection{Main Contributions}

The main contributions of this paper are as follows.

\medskip\noindent
\textbf{1. Individual-sequence Fano inequality
(Theorem~\ref{thm:fano}).}
Let $x^n$ be a channel input 
sequence that produces a channel output $y^n$,
and let $\hat{x}^n$ be the reconstruction of $x^n$ produced by a
block finite-state decoder of block length $k$ (to be defined precisely).
Let $\hat{P}_e = \frac{1}{n}\sum_{t=1}^n
\mathbf{1}\{\hat{x}_t \neq x_t\}$ denote the
fraction of incorrectly decoded symbols of $x^n$.
Then,
\begin{equation}
\label{eq:fano-preview}
    \phi(\hat{P}_e) \geq
    \bar{\rho}_k(x^n|y^n) - \Gamma_0,
\end{equation}
where $\bar{\rho}_k(x^n|y^n)$ is the block-average
conditional LZ complexity, namely, the average compression ratio of conditional LZ compression \cite{Ziv85} of
$x^n$ in blocks of length $k$
using $y^n$ as side information (to be defined precisely in the sequel),
$\Gamma_0 \to 0$ as $n\to\infty$ for fixed $k$ and a fixed number of states
of the decoder.
We believe that this individual-sequence analogue of Fano's inequality is interesting in its own right as a universal lower
bound on empirical error frequency, and we will have more to say about it
later.

\medskip\noindent
\textbf{2. Channel coding converse
(Theorem~\ref{thm:channel-tier2}).}
The aforementioned Fano inequality sets the stage for a converse to a channel
coding theorem that applies to
a completely arbitrary channel.
The channel model is $y^n = q(x^n, v^n)$ for an arbitrary
deterministic function $q$, with $v^n \in \mathcal{N}$
a fixed but unknown individual noise sequence.
As mentioned above, the decoder is assumed to be a block finite-state machine (block-FSM); 
Under these conditions, define the
\emph{individual-sequence LZ channel capacity}
$C_k(\mathcal{N})$ as the maximin LZ mutual information over input $k$-block types
and noise sequences in $\mathcal{N}$
(see Definition~\ref{def:capacity-k}).
Then, for any rate-$R$ channel code,
\begin{equation}
\label{eq:converse-preview}
    R \leq C_k(\mathcal{N}) + \phi(\hat{P}_e) + \Gamma_0.
\end{equation}
This is the individual-sequence analogue of
the classical converse $R \leq C + \phi(P_e)$.

\medskip\noindent
\textbf{3. Achievability theorem
(Theorem~\ref{thm:achievability}).}
For the class of \emph{symbol-wise invertible} channels
$y_t = f(x_t,v_t)$ (Definition~\ref{def:sym-inv}) with a
fixed but unknown individual noise sequence $v^n \in \mathcal{N}$,
and for noise sets $\mathcal{N}$ that are closed under $k$-block
permutations,
we prove that any rate strictly below $C_k(\mathcal{N})$
(Definition~\ref{def:capacity-k}) is achievable using a universal decoder
that requires no knowledge of the channel and without feedback.
The achievability is proved via a random coding argument yielding
a \emph{randomized} code (with shared randomness between encoder and decoder)
whose expected per-message error frequency vanishes simultaneously for
every $v^n \in \mathcal{N}$.
The converse (Theorem~\ref{thm:channel-tier2}), on the other hand, is proved for deterministic
codes, so there is some degree of mismatch between the two. This will be
discussed later,
in Remark~\ref{rem:gap} after Theorem~\ref{thm:achieve-emp}).

Together, results 1--3 constitute an individual-sequence
analogue of the classical channel coding theorem:
reliable transmission is possible if and only if
the code rate does not exceed the LZ channel capacity.
The individual-sequence framework for channel coding
relates to joint source-channel coding studied
in~\cite{Ziv80,me14,me22}, though in those works the
channel is still probabilistic (memoryless), whereas
here the channel noise is treated as an individual sequence.

\medskip\noindent
\textbf{Fixed noise versus worst-case guarantee.}
The noise sequence $v^n$ is an individual, fixed sequence
that is committed before any transmission takes place ---
in particular, before the encoder selects a codeword.
However, neither the encoder nor the decoder knows which
element of $\mathcal{N}$ was selected; both know only the
set $\mathcal{N}$.
The code must therefore be designed to work reliably for
every possible $v^n \in \mathcal{N}$ simultaneously.
For this reason, the converse and achievability are stated in
terms of the worst-case noise in $\mathcal{N}$: not because
$v^n$ varies during transmission, but because the code is
chosen in ignorance of which fixed $v^n$ will occur.
The fixed-in-advance nature of $v^n$ corresponds to an
\emph{oblivious} adversary: since $v^n$ is committed before
the codeword is transmitted, the adversary cannot adapt the
noise to the actual transmitted sequence.
It is the encoder's ignorance of which $v^n \in \mathcal{N}$
was committed --- not any adaptive power of the adversary ---
that forces the maximin structure of the capacity: the
encoder optimizes the codebook first, against the worst-case
noise in $\mathcal{N}$.

\subsection{Related Work}

The individual-sequence approach to source-channel coding
was initiated by Ziv~\cite{Ziv80}, who established a
separation theorem for individual source sequences
transmitted over a probabilistic discrete memoryless channel (DMC).
This was extended and further developed in~\cite{me14,me22}.
The source coding theorems of~\cite{Ziv78} are also
foundational to this line of work.
The conditional LZ complexity~\cite{Ziv85} and its
application to universal decoding over finite-state
channels~\cite{me00} are the direct predecessors of this
work.

The most closely related body of work is due to Feder and
his collaborators~\cite{SF11,LF14,LF11,LF13b}.
Shayevitz and Feder~\cite{SF11} and Lomnitz and
Feder~\cite{LF14,LF11,LF13b} study individual-sequence channel
coding for the \emph{modulo-additive} channel
$y_t = x_t \oplus v_t$, where $\oplus$ is group addition,
$x_t$ is the channel input at time $t$, $y_t$ is the output,
and $v^n$ is a fixed but unknown individual noise sequence.
They characterize the capacity as the logarithm of the output
alphabet size (in bits) minus the LZ complexity of the noise, and
construct achievability schemes, some requiring feedback
(see point~(iii) below for details).

The present paper differs from this line of work in
several respects.

\medskip\noindent
\emph{(i) General channel model.}
Our channel model $y^n = q(x^n, v^n)$ strictly generalizes
the modulo-additive model $y^n = x^n \oplus v^n$ of Feder
{\em et al.}, and their results can be recovered as a special case
of ours (see Section~\ref{sec:capacity}).
In particular, our adversary model is compatible with theirs:
in all these works, the noise sequence $v^n$ is committed
before the codeword is transmitted (an \emph{oblivious}
adversary in the terminology of Langberg~\cite{Langberg04}),
and does not depend on the codebook or the transmitted sequence.

\medskip\noindent
\emph{(ii) Fano inequality and converse.}
Shayevitz and Feder~\cite{SF11} prove achievability only
and explicitly leave the converse as an open problem.
Lomnitz and Feder~\cite{LF14,LF11,LF13b} prove a converse via
a counting argument: fixing $y^n$ and showing that too many
codewords have high empirical mutual information with $y^n$
forces a decoding error.
This yields the qualitative conclusion that $R > C_k$
implies $\hat{P}_e$ is bounded away from zero, but does not
give the quantitative form $R \leq C_k + \phi(\hat{P}_e)$.
To our knowledge, no individual-sequence analogue of Fano's
inequality appears in the prior literature; our Fano inequality
is what enables the quantitative converse.

\medskip\noindent
\emph{(iii) Feedback and common randomness.}
The achievability schemes of~\cite{SF11,LF14} require feedback
from the decoder to the encoder, while the converse
of~\cite{LF14,LF11} allows no feedback.
This creates a gap in their framework: since feedback can
increase capacity for channels with memory (as discussed
in Section~\ref{sec:discussion}), their achievability may
be operating above the non-feedback capacity, and the
converse of~\cite{LF14,LF11} does not bound the feedback capacity.
Our achievability (Theorems~\ref{thm:achieve-emp} and~\ref{thm:achievability})
requires no feedback. However, as noted above, our achievability proof
refers to a \emph{randomized} code, while the converse applies to
\emph{deterministic} codes.
Both frameworks therefore exhibit gaps between their direct and converse theorems,
but we argue that the gap in our case is somewhat softer.
See Remark~\ref{rem:gap} for a detailed comparison.

\medskip\noindent
\emph{(iv) Relation to the general probabilistic capacity formula.}
Verd\'{u} and Han~\cite{VH94} established a general formula
for the capacity of any channel (with no assumptions of
stationarity, ergodicity, or memorylessness) in terms of
the limsup in probability of the normalized information
density $\frac{1}{n}\log\frac{dP_{Y^n|X^n}}{dP_{Y^n}}$.
Their setting is probabilistic but otherwise assumption-free
on the channel side, which is the closest classical analogue
of our converse, where the channel is likewise an arbitrary
deterministic mapping with no structural assumptions.
The normalized information density of Verd\'{u}-Han is the
probabilistic counterpart of our individual-sequence LZ
mutual information: both measure the
reduction in uncertainty about the output when the input is
known, one in a distributional sense and the other in a
compressibility sense.
The key difference is that Verd\'{u}-Han work with a
channel distribution and characterize capacity as a
probabilistic limit, while here there are no distributions
anywhere: the channel is a fixed deterministic mapping, and
the bound holds for every individual sequence pair
$(x^n, y^n)$ without any probabilistic averaging.
Our converse also requires a finite-state decoder, a
condition that has no counterpart in the Verd\'{u}-Han
framework.

\medskip\noindent
\emph{(v) Relation to the classical AVC literature.}
The arbitrarily varying channel (AVC) is a classical model
in which the channel is selected, for each transmitted symbol
or block, by an adversary from a known family.
Blackwell, Breiman, and Thomasian~\cite{BBT60} established
the random coding capacity of the AVC, showing it equals
the capacity of the compound channel formed by the worst-case
member of the family.
Ahlswede~\cite{Ahlswede78} proved the fundamental \emph{AVC dichotomy}:
under the average-error criterion, the deterministic code capacity
is either zero or equals the random coding capacity.
The boundary between the two regimes is characterized by
\emph{symmetrizability}: Csisz\'{a}r and Narayan~\cite{CN88}
showed that the deterministic capacity equals the random coding
capacity if and only if the AVC is not symmetrizable.
Hughes and Thomas~\cite{HT96} and
Csisz\'{a}r and Narayan~\cite{CN91}
extended these results to AVCs with input and state cost constraints.
Langberg~\cite{Langberg04} studied the distinction between
oblivious and non-oblivious adversaries, showing that
obliviousness strictly limits the adversary's power.

The present paper departs from the AVC framework in two
fundamental respects.
First, in the classical AVC the adversary selects the
channel state according to some strategy from a \emph{set of
distributions}; here the noise $v^n$ is a completely
individual deterministic sequence with \emph{no distribution
assumed anywhere} --- not even the existence of a probability
space is required.
Second, and more importantly, the converse is established
via an \emph{individual-sequence Fano inequality} (Theorem~\ref{thm:main-tier2}),
a new tool with no direct counterpart in the AVC literature
(where converses rely on probabilistic arguments such as
the blowing-up lemma or the method of types).
Symmetrizability plays no role in our setting: the
maximin capacity $C_k(\mathcal{N})$ is well-defined
and positive whenever the channel is symbol-wise invertible
and the noise budget $\rho_v$ is smaller than the channel input
entropy, with no symmetrizability condition needed.

Eswaran, Sarwate, Sahai, and Gastpar~\cite{ESSG10} study
an AVC driven by an individual state sequence $z^n$: for each
state value $z \in \mathcal{Z}$, the channel is a DMC $W(y|x,z)$,
and the state sequence is arbitrary and unknown.
They show that zero-rate feedback combined with common randomness
suffices to achieve rates approaching the empirical mutual
information $I(P, W_{z^n})$, where $W_{z^n}$ is the
state-averaged channel.
The present paper differs from~\cite{ESSG10} in three respects.
First, our channel model is strictly more general: we impose no
probabilistic transition structure --- the channel is a fully
deterministic mapping $x^n \mapsto y^n$ with an individual noise
sequence, whereas their channel is a DMC for each state value.
Second, we require no feedback: common randomness alone suffices,
which strictly dominates their requirement of zero-rate feedback
plus common randomness.
Third, we provide a matching converse (Theorem~\ref{thm:channel-tier2}),
making ours a complete coding theorem; \cite{ESSG10} establishes
only an achievability result.

\subsection{Paper Organization}

The paper is structured around a two-tier proof strategy.
Rather than working directly at the LZ complexity level, we first establish 
a complete coding theorem --- converse and achievability --- at the level of
\emph{empirical entropies}, 
a more tractable complexity measure defined by the empirical distribution of
non-overlapping blocks of a fixed order.
The converse is then lifted to the LZ level via an upper sandwiching inequality
that relates block-average conditional LZ complexity to empirical conditional entropy;
the achievability is proved at both levels --- first at the empirical level
(Theorem~\ref{thm:achieve-emp}), and then directly at the LZ level.
This two-tier structure is the reason Section~\ref{sec:capacity}
(empirical-entropy coding theorem) precedes Section~\ref{sec:lz-capacity}
(LZ coding theorem); the latter builds directly on the former.

Section~\ref{sec:model} establishes the model,
notation, LZ complexity background, and the upper sandwiching
bound of~\cite{p246}.
Section~\ref{sec:fano} proves the individual-sequence
Fano inequality (Theorems~\ref{thm:fano}
and~\ref{thm:main-tier2}) and its generalization to
difference distortion measures
(Theorem~\ref{thm:fano-distortion}).
Section~\ref{sec:capacity} establishes the coding theorem
at the empirical-entropy level: it defines the empirical
capacity $\hat{C}_m(\mathcal{N})$ for general block-permutation-invariant noise sets
$\mathcal{N}$ (Definition~\ref{def:capacity}), proves the Fano-based
converse (Corollary~\ref{cor:capacity-tier1}), and proves
achievability at the empirical-entropy level
(Theorem~\ref{thm:achieve-emp}), with specialization to
the empirical entropy ball $\mathcal{N}_{\mathrm{emp}}(H_v)$ in Remark~\ref{rem:closed-form}.
Section~\ref{sec:lz-capacity} lifts these results to the
LZ level: it defines $C_k(\mathcal{N})$
(Definition~\ref{def:capacity-k}), proves the LZ converse
(Theorem~\ref{thm:channel-tier2}), and proves LZ
achievability (Theorem~\ref{thm:achievability}).
Section~\ref{sec:lz-capacity} also contains two subsections with additional results:
Section~\ref{sec:jscc} applies the full coding theorem to joint
source-channel coding, establishing an individual-sequence
source-channel separation theorem at both the empirical-entropy
level (Lemma~\ref{lem:equiv}, Theorem~\ref{thm:channel-tier1-jscc})
and the LZ complexity level (Remark~\ref{rem:jscc-lz});
Section~\ref{sec:discussion} discusses open problems.

\section{Model, Notation, and Background}
\label{sec:model}

\subsection{Channel Model and Problem Formulation}

We consider a communication system in which a channel input
sequence $x^n \in \calX^n$ is transmitted over a channel,
and the channel output is $y^n\in\calY^n$.
The alphabets $\calX$, $\calY$ are finite with
sizes $\alpha$, $\beta$, respectively.
The noise alphabet $\calV$ has size $\nu$.
All logarithms are base~2.

\medskip\noindent
\textbf{Notation.}
Given $x^n$ and integers
$1 \leq i \leq j \leq n$, we use $x_i^j$ to denote the
segment $(x_i, x_{i+1}, \ldots, x_j)$. When $i=1$, the subscript
is omitted, thus $x_1^n = x^n$.
The same convention applies to $y^n$ and $v^n$.
Calligraphic letters ($\calX$, $\calY$, $\calV$)
denote the finite alphabets introduced above.
The indicator function of an event $\calA$ is written
$\mathbf{1}\{\calA\}$, and $h_2(u) \dfn -u\log u - (1-u)\log(1-u)$,
$u\in[0,1]$,
denotes the binary entropy function.

\medskip\noindent
\textbf{Encoder.}
For the converse, the encoder is an \emph{arbitrary} mapping
that maps a message into a channel input vector, $x^n\in \calX^n$.
For the achievability, the encoder is defined by a codebook
(see Definition~\ref{def:ensemble}).

\medskip\noindent
\textbf{Channel model (converse).}
For the converse, the channel is a fully general deterministic
function $q: \calX^n \times \calV^n \to \calY^n$, with a fixed
but unknown individual noise sequence $v^n \in \mathcal{N}$:
\begin{equation}
\label{eq:general-channel}
    y^n = q(x^n, v^n),
\end{equation}
with no structural assumptions on $q$ beyond determinism.

\medskip\noindent
\textbf{The role of the noise set $\mathcal{N}$.}
The noise sequence $v^n$ is an individual, fixed sequence 
that does not depend on the
codebook, the message, or the transmitted codeword.
The noise set $\mathcal{N} \subseteq \mathcal{V}^n$ is a design
parameter known to both encoder and decoder: it specifies the
class of noise sequences against which the code is required to
be reliable.

This model is non-trivial despite $v^n$ being fixed, because
the decoder is constrained to a block-FSM with $s$ states of
inter-block memory (see below). Had the decoder been unconstrained,
then for fixed $v^n$, the optimal reference decoder could have canceled the noise
and thereby recovered the message $x^n$ (at least under many 
channel models, like the modulo-additive channel), rendering the problem trivial and
uninteresting. The block-FSM constraint prevents this: for large $n$, the decoder cannot
store $v^n$ across blocks or exploit its structure globally.
It is this computational limitation on the decoder, not
uncertainty about $v^n$, that makes reliable communication
over the channel a genuinely non-trivial problem.
The maximin capacity formula~\eqref{eq:capacity-k} reflects
this structure: the maximum over $x^n$ represents the
encoder's commitment based on $\mathcal{N}$ alone, and the
minimum over $v^n \in \mathcal{N}$ is the worst-case guarantee
the code must meet simultaneously for every possible
realization of the fixed noise sequence.
In the language of the AVC literature (Section~\ref{sec:intro},
point~(v)), this corresponds to an \emph{oblivious} adversary:
$v^n$ is committed before the codeword is transmitted and
does not depend on it. As mentioned before,
this is also the model of Shayevitz and Feder~\cite{SF11}
and Lomnitz and Feder~\cite{LF14,LF11,LF13b}, with whom
our adversarial assumptions are compatible.

\medskip\noindent
\textbf{Channel model (achievability).}
For the achievability, two conditions are required.
First, the channel must be \emph{symbol-wise invertible}
(Definition~\ref{def:sym-inv} below): the output at each
time $t$ depends only on the current input and noise symbols,
and the noise can be recovered from input and output.
Second, the noise set $\calN$ must be \emph{closed under
$k$-block permutations}: for every $v^n \in \calN$ and every
permutation $\pi$ of the $n/k$ consecutive $k$-blocks of
$v^n$, the permuted sequence $v^n_\pi$ also belongs to
$\calN$ (this is satisfied by the LZ ball
$\calN_{\mathrm{LZ}}(\rho_v)$ and by Hamming balls).

\begin{definition}[Symbol-wise invertible channel]
\label{def:sym-inv}
A deterministic channel is \emph{symbol-wise invertible}
if there exists a function $f:\calX\times\calV\to\calY$
such that $y_t = f(x_t, v_t)$ for every $t = 1,\ldots,n$,
and for every fixed $v\in\calV$ the map
$f(\cdot,v):\calX\to\calY$ is a bijection.
\end{definition}

\noindent
This class is broader than it may first appear.
Three families of practical interest are:

\begin{enumerate}
\item \textbf{Group-additive channels.}
$y_t = x_t \oplus v_t$, where $\calX = \calY$ is a finite abelian
group and $\oplus$ is the group operation.
For every fixed $v$, the map $x \mapsto x \oplus v$ is a
group translation, hence a bijection.
This is the model studied in~\cite{SF11,LF14,LF11}.

\item \textbf{Affine channels over a finite field.}
$\calX = \calY = \mathbb{F}_q$, $y_t = a \cdot x_t + v_t$ for a fixed
nonzero field element $a \in \mathbb{F}_q^*$ and noise $v_t \in \mathbb{F}_q$.
For every fixed $v$, the map $x \mapsto ax + v$ is an affine bijection.
This subsumes the group-additive case ($a = 1$) and captures
channels with a multiplicative gain.

\item \textbf{Permutation channels.}
$\calX = \calY$, $y_t = \pi_{v_t}(x_t)$, where
$\{\pi_v\}_{v \in \calV}$ is an arbitrary family of permutations
of $\calX$ (not necessarily forming a group or even a coset).
The noise symbol $v_t$ selects which permutation is applied at
time $t$, independently of all others.
This is strictly more general than group-additive channels,
since the permutations $\pi_v$ need not be group translations.
\end{enumerate}

\medskip\noindent
\textbf{Decoder (converse vs.\ achievability).}
A \emph{decoder} maps the channel output sequence
$y^n\in\calY^n$
to an estimate $\hat{x}^n
\in \calX^n$ of the transmitted channel input sequence.
For the \emph{converse}, the decoder is a
\emph{block finite-state machine (block-FSM)} with $s$ states and
block length $m$.
At each block step $i = 0,1,\ldots,n/m-1$, it implements the pair of
equations:
\begin{equation}
\label{eq:fs-decoder}
    \hat{x}_{im+1}^{(i+1)m} = f(z_i,y_{im+1}^{(i+1)m}),
    \qquad
    z_{i+1}= g(z_i,y_{im+1}^{(i+1)m}),
\end{equation}
where $z_0,z_1, z_2,\ldots$ take values in a finite set
$\calZ$ of size $s$, and $z_0$ is fixed.

For the \emph{achievability}, the decoder is the $k$-block
resetting conditional LZ decoder (Definition~\ref{def:decoder}),
which selects the codeword minimizing the block-average conditional
LZ complexity and requires
no knowledge of the channel. 

\medskip\noindent
\textbf{Decoder state count.}
The converse decoder (block length $m$) gives a state-count correction term
of the form $(\log s)/m$, which vanishes as $m \to \infty$ for any fixed $s$.
The achievability decoder (block length $k$) has $s = 1$, so its correction
term vanishes identically.

The block-FSM model strictly generalizes the classical
symbol-by-symbol FSM: any ordinary FSM run for $m$ steps
can be viewed as an $m$-block FSM (by taking the $m$-step
product of its transition and output functions), but the
converse is false --- a block-FSM output function
$f: \calZ\times \calY^m \to \calX^m$
can implement arbitrary block-to-block mappings that no
symbol-by-symbol FSM with the same state space can realize.
At first glance, the condition $\log s \ll m$ might appear to limit the converse's
generality, since a machine that accumulates memory
over past blocks would need state cardinality exponential in $m$,
making $(\log s)/m = O(1)$ and the correction term non-vanishing.
The key point is that the condition is not a restriction
on the decoder's computational power \emph{within} a block --- the
output function $f$ may be arbitrarily
complex --- but solely on the amount of \emph{inter-block memory}.

\medskip\noindent
\textbf{Error measure and key function.}
The \emph{empirical symbol-error frequency} is:
\begin{equation}
\label{eq:pe-def}
    \hat{P}_e \dfn \frac{1}{n}\sum_{t=1}^n
    \mathbf{1}\{\hat{x}_t \neq x_t\}.
\end{equation}
The function $\phi$, which plays the role of the
right-hand side of Fano's
inequality~\eqref{eq:classical-fano}, is:
\begin{equation}
\label{eq:phi-def}
    \phi(p) \dfn h_2(p) + p\log(|\calX|-1).
\end{equation}
It is strictly increasing on $[0,1-1/|\calX|]$, with
maximum $\log\alpha$ and hence invertible in this range, with inverse
denoted $\phi^{-1}:[0,\log|\calX|]\to[0,1-1/|\calX|]$.

\subsection{Normalized Empirical Entropies}

Proofs use normalized empirical entropies as intermediate
quantities.
For positive integer $m$ dividing $n$, let
$\hat{P}(x^m,y^m) = \frac{m}{n}\sum_{i=0}^{n/m-1}
\mathbf{1}\{x_{im+1}^{im+m}=x^m\}\cdot
\mathbf{1}\{y_{im+1}^{im+m}=y^m\}$
and define
\begin{equation}
\label{eq:hHm-def}
    \hat{H}_m(x^n) \dfn
    -\frac{1}{m}\sum_{x^m}
    \hat{P}(x^m)\log\hat{P}(x^m).
\end{equation}
The conditional version is $\hat{H}_m(x^n|y^n) \dfn
-\frac{1}{m}\sum_{x^m,y^m}\hat{P}(x^m,y^m)
\log\hat{P}(x^m|y^m)$, and the mutual
information is $\hat{I}_m(x^n;y^n) \dfn
\hat{H}_m(x^n) - \hat{H}_m(x^n|y^n)$.
Sometimes a more convenient notation would be the following:
let $X^m$ and $Y^m$ be an auxiliary random $m$-vectors jointly distributed
according to $\hat{P}$.
Then $\hat{H}_m(x^n) = \hat{H}(X^m)/m$ and
$\hat{H}_m(x^n|y^n) = \hat{H}(X^m|Y^m)/m$,
where $\hat{H}(\cdot)$ and $\hat{H}(\cdot|\cdot)$ denote respectively
the Shannon entropy and conditional entropy of the auxiliary
random vectors under $\hat{P}$. Similar notation rules will be applied also 
for other random vectors.

\section{An Individual-Sequence Analogue of Fano's Inequality}
\label{sec:fano}

\subsection{Basic Form}

\begin{theorem}[Individual-sequence Fano inequality, basic form]
\label{thm:fano}
Given $x^n, \hat{x}^n \in \calX^n$ and a positive
integer $m$ dividing $n$:
\begin{equation}
\label{eq:fano-tier1}
    \hHm{m}(x^n|\hat{x}^n) \leq \phi(\hat{P}_e).
\end{equation}
\end{theorem}

\begin{proof}
Let
\begin{equation}
\label{eq:Pe-j}
    \hat{P}_e[j] \dfn \frac{m}{n}\sum_{i=0}^{n/m-1}
    \mathbf{1}\{x_{im+j} \neq \hat{x}_{im+j}\},
    \quad j=1,\ldots,m,
\end{equation}
so that $\frac{1}{m}\sum_{j=1}^m \hat{P}_e[j] = \hat{P}_e$.
Using the auxiliary-vector notation, 
the chain rule and the fact that conditioning reduces entropy yield
\begin{equation}
	\hHm{m}(x^n|\hat{x}^n)=\frac{\hat{H}(X^m|\hat{X}^m)}{m}
    = \frac{1}{m}\sum_{j=1}^m \hat{H}(X_j|X_1^{j-1},\hat{X}^m)
    \leq \frac{1}{m}\sum_{j=1}^m \hat{H}(X_j|\hat{X}_j).
\end{equation}
Applying the classical Fano inequality to each term:
\begin{eqnarray}
\hHm{m}(x^n|\hat{x}^n)&\le& 
\frac{1}{m}\sum_{j=1}^m \hat{H}(X_j|\hat{X}_j)\nonumber\\
&\leq& \frac{1}{m}\sum_{j=1}^m\phi(\hat{P}_e[j])\nonumber\\
&\leq& \phi\left(\frac{1}{m}\sum_{j=1}^m\hat{P}_e[j]\right)\nonumber\\
&=&\phi(\hat{P}_e), \qedhere
\end{eqnarray}
where the last inequality is by the concavity of $\phi(\cdot)$.
\end{proof}

\subsection{Fano's Inequality in Terms of the Channel Output}
\label{sec:main}

\begin{theorem}[Individual-sequence Fano inequality, output form]
\label{thm:main-tier2}
Let $x^n \in \calX^n$, $y^n \in \calY^n$ be given
and let $\hat{x}^n$ be the output of a
block-FSM decoder with block length $m$ (dividing $n$) and $s$ states,
applied to $y^n$.
Then:
\begin{equation}
\label{eq:main-tier2}
    \phi(\hat{P}_e) \geq \hHm{m}(x^n|y^n) - \frac{\log s}{m}.
\end{equation}
\end{theorem}
For $m \gg \log s$, the term $(\log s)/m$ is small
and the main term remains $\hHm{m}(x^n|y^n)$.

\begin{proof}
At step $i$ the decoder
reads $y_{im+1}^{(i+1)m}$, produces
$\hat{x}_{im+1}^{(i+1)m} = f(z_i, y_{im+1}^{(i+1)m})$,
and updates $z_{i+1} = g(z_i, y_{im+1}^{(i+1)m})$.
Define
\begin{equation}
\label{eq:joint-emp}
    \hat{P}(z, x^m, y^m, \hat{x}^m)\dfn
    \frac{m}{n} \sum_{i=0}^{n/m-1}
    \mathbf{1}\bigl\{
        z_i = z,\;
        x_{im+1}^{(i+1)m} = x^m,\;
        y_{im+1}^{(i+1)m} = y^m,\;
        \hat{x}_{im+1}^{(i+1)m} = \hat{x}^m
    \bigr\}.
\end{equation}
Let $(Z, X^m, Y^m, \hat{X}^m)$ denote the corresponding random
variables under $\hat{P}$ using the alternative notation rules.
By the relation $\hat{X}^m = f(Z,Y^m)$
\begin{align}
    \phi(\hat{P}_e)&\geq \frac{\hat{H}(X^m|\hat{X}^m)}{m}\\
    &\geq \frac{\hat{H}(X^m|\hat{X}^m, Z, Y^m)}{m}
    \label{eq:dp1}\\
    &= \frac{\hat{H}(X^m|Z, Y^m)}{m}
    \label{eq:dp2}\\
    &= \frac{\hat{H}(X^m|Y^m)}{m} - \frac{\hat{I}(Z;\, X^m \mid Y^m)}{m}\\
    &\ge \frac{\hat{H}(X^m|Y^m)}{m} - \frac{\hat{H}(Z|Y^m)}{m}\\
    &\ge \frac{\hat{H}(X^m|Y^m)}{m} - \frac{\log s}{m}\\
    &= \hHm{m}(x^n|y^n) - \frac{\log s}{m}.
    \label{eq:dp3}
\end{align}
\end{proof}

\subsection{Generalized Fano Inequality for Difference Distortion Measures}
\label{sec:fano-distortion}

The individual-sequence Fano inequality established above is 
interesting in its own right, beyond its role in the channel coding converse.
Before proceeding to the converse, we pause to examine it 
from a somewhat broader perspective by generalizing it to the setting 
of general difference distortion measures.
This generalization shows that the Fano inequality captures 
a structural property of empirical entropy that goes beyond the channel coding context.

The Hamming distortion frequency, $\hat{P}_e$, is a natural distortion
measure when $\calX$ is an unstructured alphabet, but in many
settings the alphabet members carry quantitative values --- for
example, $\calX = \mathbb{Z}_m$ with a meaningful notion of
distance between symbols.
In such cases a finer distortion measure, one that penalizes
large errors more than small ones, is more informative than
the raw fraction of incorrect symbols.
We now extend Theorem~\ref{thm:main-tier2} to a general
\emph{difference distortion measure}.
Suppose $\calX$ is a finite abelian group and
$\Delta: \calX \to [0,\infty)$ satisfies $\Delta(0)=0$.
The distortion between $x$ and $\hat{x}$ is defined as $\Delta(x - \hat{x})$.
The empirical average distortion is
\begin{equation}
\label{eq:empirical-distortion}
    \bar{\Delta}(x^n,\hat{x}^n)
    \dfn \frac{1}{n}\sum_{t=1}^n \Delta(x_t - \hat{x}_t).
\end{equation}
Define the \emph{maximum-entropy function}
\begin{equation}
\label{eq:Phi-def}
    \Phi(D) \dfn \max\bigl\{H(Z) : Z \text{ r.v.\ on } \calX,\;
    \mathbb{E}[\Delta(Z)] \leq D\bigr\},
\end{equation}
which is concave and non-decreasing in $D \geq 0$, with $\Phi(0) = 0$.

\begin{theorem}[Generalized Fano inequality for difference distortion]
\label{thm:fano-distortion}
Under the same hypotheses as Theorem~\ref{thm:main-tier2}:
\begin{equation}
\label{eq:fano-distortion}
    \Phi(\bar{\Delta}(x^n,\hat{x}^n))
    \geq \hHm{m}(x^n|y^n) - \frac{\log s}{m}.
\end{equation}
\end{theorem}

\begin{proof}
Write $e_t \dfn x_t - \hat{x}_t$ for the error symbol at time $t$,
so $\bar{\Delta}(x^n,\hat{x}^n) = \frac{1}{n}\sum_t \Delta(e_t)$.
Since $x_t = \hat{x}_t + e_t$, knowing $\hat{x}^n$ reduces
estimating $x^n$ to estimating $e^n$:
\begin{equation}
    \hHm{m}(x^n|\hat{x}^n) = \hHm{m}(e^n|\hat{x}^n)
    \leq \hHm{m}(e^n) \leq \hat{H}_1(e^n),
\end{equation}
where: the first step is an equality because $x^n \mapsto e^n = x^n - \hat{x}^n$
is a bijection given $\hat{x}^n$;
the second inequality is data-processing (dropping conditioning);
the third uses $\hat{H}_m \leq \hat{H}_1$, which follows by the facts
that conditioning reduces entropy and
concavity of the entropy.
Since $\frac{1}{n}\sum_t \Delta(e_t)=\frac{1}{n}\sum_t \Delta(x_t-\hat{x}_t)$,
by the definition of $\Phi(\cdot)$,
\begin{equation}
\label{eq:dist-step1}
    \hHm{m}(x^n|\hat{x}^n)\le\hat{H}_1(e^n)
    \leq \Phi\bigl(\bar{\Delta}(x^n,\hat{x}^n)\bigr).
\end{equation}
But from the proof of
Theorem~\ref{thm:main-tier2},
\begin{equation}
    \hHm{m}(x^n|\hat{x}^n) \geq \hHm{m}(x^n|y^n) - \frac{\log s}{m},
\end{equation}
and so,
\begin{equation}
\label{eq:dist-Hky-bound}
    \hHm{m}(x^n|y^n) \leq \Phi\bigl(\bar{\Delta}(x^n,\hat{x}^n)\bigr) + \frac{\log s}{m}.
\end{equation}
\end{proof}

\section{Empirical Channel Capacity}
\label{sec:capacity}

\subsection{Capacity Definition and Converse}

\begin{definition}[Empirical channel mutual information
and capacity]
\label{def:capacity}
For channel input $x^n$, output $y^n$, and block length $m$:
\begin{equation}
\label{eq:channel-MI}
    \hat{I}_m(x^n;y^n) \dfn
    \hHm{m}(x^n) - \hHm{m}(x^n|y^n).
\end{equation}
For a given noise set $\mathcal{N}$, the
\emph{empirical channel capacity of order $m$} is:
\begin{equation}
\label{eq:capacity}
    \hat{C}_m(\mathcal{N}) \dfn
    \max_{x^n \in \calX^n}
    \min_{v^n \in \mathcal{N}}
    \hat{I}_m(x^n;y^n),
\end{equation}
where $y^n = q(x^n,v^n)$ is the channel output~\eqref{eq:general-channel}.
\eoe
\end{definition}

This is the maximin of the empirical mutual information
over inputs and noise sequences in $\mathcal{N}$.
The per-realization capacity $C_m(v^n) \dfn
\max_{x^n}\hat{I}_m(x^n;y^n)$ for the specific
$v^n$ satisfies
$C_m(v^n) \geq \hat{C}_m(\mathcal{N})$ for every
$v^n \in \mathcal{N}$.

\begin{remark}[Input constraints]
\label{rem:input-constraints}
Input constraints are handled by restricting the maximum
in~\eqref{eq:capacity} to a set
$\mathcal{A} \subseteq \calX^n$:
\begin{equation}
\label{eq:capacity-constrained}
    \hat{C}_m(\mathcal{A}, \mathcal{N}) \dfn
    \max_{x^n \in \mathcal{A}}
    \min_{v^n \in \mathcal{N}}
    \hat{I}_m(x^n;y^n).
\end{equation}
The converse (Corollary~\ref{cor:capacity-tier1}) extends
to the constrained case with the same proof, since
Theorem~\ref{thm:main-tier2} holds for any $x^n$
regardless of $\mathcal{A}$.
The LZ-level extension is discussed in Section~\ref{sec:lz-capacity}.

For the achievability, the constraint must be compatible
with the probabilistic method: the codebook is drawn from
the appropriate type class of the capacity-achieving $x^n$
(first-order for memoryless constraints, $r$-th order for
memory-$r$ constraints, or the finite-state type of
Cs\'{i}sz\'{a}r~\cite{Csiszar98} for automaton-defined
constraints such as $(d,k)$ RLL codes).
In every case the type class size is exponential in $n$
(with at most a polynomial correction factor), so the
achievability proof goes through with the polynomial
prefactor replacing $(n+1)^\alpha$.
\eoe
\end{remark}

\begin{theorem}[Converse Theorem in empirical-entropy terms]
\label{cor:capacity-tier1}
Under the conditions of Theorem~\ref{thm:main-tier2},
suppose a codebook $\mathcal{C}_n$ of $M = 2^{nR}$ codewords achieves
\begin{equation}
\label{eq:Pe-codebook}
    \max_{x^n \in \mathcal{C}_n}\;\max_{v^n \in \mathcal{N}}\;
    \frac{1}{n}\sum_{t=1}^n \mathbf{1}\{\hat{x}_t(q(x^n,v^n)) \neq x_t\}
    \;\leq\; \delta,
\end{equation}
i.e., the worst-case symbol-error frequency over all codewords and all $v^n \in \mathcal{N}$
is at most $\delta$.
Then:
\begin{equation}
\label{eq:capacity-bound-tier1}
    R \leq \hat{C}_m(\mathcal{N})
    + \phi(\delta)
    + \frac{\log s}{m} + o(1),
\end{equation}
where $o(1) \to 0$ as $n \to \infty$ for fixed $m$.
\end{theorem}

\begin{proof}
Any codebook of $M = 2^{nR}$ distinct codewords over $\calX^n$
has at most $(n/m+1)^{\alpha^m}$ distinct order-$m$ empirical types.
By the pigeonhole principle, there exists a type $P_{X^m}$ such that
at least $2^{nR - o(n)}$ codewords share type $P_{X^m}$.
Since all these codewords are distinct elements of the type class
$\calT_m(P_{X^m})$, and $|\calT_m(P_{X^m})| \leq 2^{nH(P_{X^m})/m}$
by the standard multinomial estimate, we get:
\begin{equation}
\label{eq:ct1-rate-bound}
    R \leq \frac{H(P_{X^m})}{m} + o(1).
\end{equation}
Fix this type $P_{X^m}$ and any codeword $x^n$ of this type.
Let $v_*^n \dfn \arg\min_{v^n \in \mathcal{N}}\hat{I}_m(x^n;q(x^n,v^n))$.
Since the code achieves $\hat{P}_e \leq \delta$ for every $v^n \in \mathcal{N}$,
applying Theorem~\ref{thm:main-tier2} to $(x^n, q(x^n,v_*^n))$:
\begin{align}
    \phi(\delta)\geq\phi(\hat{P}_e(v_*^n))
    &\geq \hHm{m}(x^n|y_*^n) - \frac{\log s}{m}
    \nonumber\\
    &= \hHm{m}(x^n) - \hat{I}_m(x^n;y_*^n) - \frac{\log s}{m}
    \nonumber\\
    &\geq \frac{H(P_{X^m})}{m} - \hat{C}_m(\mathcal{N})
    - \frac{\log s}{m}\nonumber\\
    &\geq R-\hat{C}_m(\mathcal{N})-\frac{\log s}{m}-o(1),
\end{align}
\qedhere
\end{proof}

\subsection{Achievability in Empirical-Entropy Terms}
\label{sec:emp-coding-thm}

Before we state and prove achievability, a few preparatory steps are in order.
The first is the following simple lemma.

\begin{lemma}
\label{lem:minimax-emp}
Let the channel be symbol-wise invertible
(Definition~\ref{def:sym-inv}).
For every $x^n \in \calX^n$ and $v^n \in \calV^n$
(with $y^n = q(x^n,v^n)$):
\begin{equation}
\label{eq:minimax-relabel}
    \hHm{m}(v^n|x^n) = \hHm{m}(y^n|x^n),
\end{equation}
and consequently:
\begin{equation}
\label{eq:minimax-decomp}
    \hat{I}_m(x^n;y^n)
    =\hHm{m}(y^n) - \hHm{m}(v^n|x^n).
\end{equation}
\end{lemma}

\begin{proof}
For fixed $x^n$, the map $v_t \mapsto y_t = f(x_t,v_t)$
is a bijection for each $x_t$ (Definition~\ref{def:sym-inv}).
Within any $m$-block, $(v^m,x^m)\mapsto(y^m,x^m)$ acts
symbol-wise as a bijection in $v\leftrightarrow y$ for fixed $x$,
so $H(\hat{P}_{V^m|X^m}) = H(\hat{P}_{Y^m|X^m})$.
Averaging over blocks gives~\eqref{eq:minimax-relabel},
and~\eqref{eq:minimax-decomp} follows from
$\hat{I}_m(x^n;y^n) = \hHm{m}(y^n) - \hHm{m}(y^n|x^n)$. \qedhere
\end{proof}

The second preparatory step is provided by the following lemma.

\begin{lemma}[Constancy of the minimum empirical mutual information]
\label{lem:ihat-constant}
Let the channel be symbol-wise invertible
(Definition~\ref{def:sym-inv}),
let $\mathcal{N} \subseteq \calV^n$ be any noise set that is closed
under $m$-block permutations (i.e., if $v^n \in \mathcal{N}$ and
$\pi$ is any permutation of the $n/m$ blocks of size $m$,
then $\pi(v^n) \in \mathcal{N}$),
and let $P_{X^m}$ be any order-$m$ type on $\calX^m$.
For every $x^n \in \mathcal{T}_m(P_{X^m})$, define
\begin{equation}
    \hat{I}_m^*(x^n)\dfn\min_{v^n \in \mathcal{N}}
\hat{I}_m(x^n; q(x^n,v^n)).
\end{equation}
Then $\hat{I}_m^*(x^n)$ takes the same value for all $x^n \in \mathcal{T}_m(P_{X^m})$;
we denote this common value $\hat{I}_m^*(\mathcal{N}, P_{X^m})$.
\end{lemma}

\begin{proof}
For any $x^n, \tilde{x}^n \in \mathcal{T}_m(P_{X^m})$,
there is an $m$-block permutation $\pi$ with $\pi(x^n) = \tilde{x}^n$.
For any $v^n \in \mathcal{N}$, set $\tilde{v}^n = \pi(v^n)$;
by the permutation-closure assumption, $\tilde{v}^n \in \mathcal{N}$ as well.
Write $\tilde{y}^n = q(\tilde{x}^n, \tilde{v}^n)$.
Since $q$ acts symbol-wise, applying $\pi$ to every block gives
$\tilde{y}^n = \pi(y^n)$ where $y^n = q(x^n,v^n)$.
The empirical mutual information depends only on the joint empirical distribution
of $m$-blocks, which is preserved by any $m$-block permutation, so
$\hat{I}_m(\tilde{x}^n; \tilde{y}^n) = \hat{I}_m(x^n; y^n)$.
The map $v^n \mapsto \tilde{v}^n = \pi(v^n)$ is a bijection on
$\mathcal{N}$, so
\begin{equation}
    \hat{I}_m^*(\tilde{x}^n)
    = \min_{\tilde{v}^n \in \mathcal{N}} \hat{I}_m(\tilde{x}^n;\tilde{y}^n)
    = \min_{v^n \in \mathcal{N}} \hat{I}_m(x^n;y^n)
    = \hat{I}_m^*(x^n).
\end{equation}
Since $x^n$ and $\tilde{x}^n$ were arbitrary elements of $\mathcal{T}_m(P_{X^m})$,
the minimum is constant on $\mathcal{T}_m(P_{X^m})$.
\end{proof}

Third, we need to define the decoder.
The decoder to be used in the achievability is the
\emph{Minimum Empirical Conditional Entropy (MECE) decoder}:
\begin{equation}
\label{eq:mece-decoder}
    \hat{w} \dfn \arg\min_{w \in \calM_n}
    \hHm{m}(x^n(w) \mid y^n),
\end{equation}
where $x^n(w)$ is the codeword pertaining to message number $w$ and
$\calM_n=\{1,2,\ldots,2^{nR}\}$.

We are now ready to assert and prove the achievability theorem.

\begin{theorem}[Achievability]
\label{thm:achieve-emp}
Let the channel be symbol-wise invertible
(Definition~\ref{def:sym-inv}),
and let $\mathcal{N} \subseteq \calV^n$ be any noise set that is closed
under $m$-block permutations.
Let $P_{X^m}$ be any order-$m$ type on $\calX^m$
(the capacity-achieving type $P^*_{X^m}$ is the natural choice, but the achievability holds for any $P_{X^m}$),
and let $\hat{I}_m^*(\mathcal{N}, P_{X^m})$ be the constant value guaranteed by
Lemma~\ref{lem:ihat-constant}.
Suppose $R < \hat{I}_m^*(\mathcal{N}, P_{X^m})$.
A randomized code of rate $R$ with $M = 2^{nR}$ codewords
drawn independently and uniformly from $\mathcal{T}_m(P_{X^m})$,
decoded by the MECE decoder~\eqref{eq:mece-decoder},
achieves:
\begin{equation}
\label{eq:ach-emp}
    \mathbb{E}_\omega\{\hat{P}_e\} \to 0
    \quad \text{as } n \to \infty,
\end{equation}
simultaneously for every $v^n \in \mathcal{N}$
and every message $w$.
Consequently, $\hat{C}_m(\mathcal{N})$ is achievable.
\end{theorem}

\begin{proof}
By Lemma~\ref{lem:ihat-constant}, $\hat{I}_m^*(\mathcal{N}, P_{X^m})$ is well-defined
as a constant on $\mathcal{T}_m(P_{X^m})$. Let
$R = \hat{I}_m^*(\mathcal{N}, P_{X^m}) - \eps$ for some $\eps > 0$.
For the transmitted codeword $x^n = x^n(w,\omega) \in \mathcal{T}_m(P_{X^m})$
and the actual $v^n \in \mathcal{N}$,
the MECE decoder assigns it score $\hHm{m}(x^n \mid q(x^n,v^n))$.
By the chain rule:
\begin{equation}
\label{eq:correct-score}
    \hHm{m}(x^n \mid y^n)
    = \hHm{m}(x^n) - \hat{I}_m(x^n;y^n)
    \leq \frac{H(P_{X^m})}{m} - \hat{I}_m^*(\mathcal{N},P_{X^m}).
\end{equation}
For fixed $y^n$, define the confusable set with the same threshold:
\begin{equation}
    \mathcal{C}(y^n) \dfn
    \bigl\{\tilde{x}^n \in \mathcal{T}_m(P_{X^m}) :
    \hHm{m}(\tilde{x}^n \mid y^n) \leq \tfrac{1}{m}H(P_{X^m}) - \hat{I}_m^*(\mathcal{N},P_{X^m})\bigr\}.
\end{equation}
By the method of types,
\begin{equation}
\label{eq:confusable-size-emp}
    |\mathcal{C}(y^n)| \leq \exp_2\{n[H(P_{X^m})/m - \hat{I}_m^*(\mathcal{N},P_{X^m})) +
o(n)]\}
\end{equation}
and
$|\mathcal{T}_m(P_{X^m})| \geq \exp_2\{nH(P_{X^m})/m\}/(n/m+1)^{\alpha^m}$.
For any wrong codeword $x^n(w',\omega)$, $w' \neq w$,
drawn uniformly from $\mathcal{T}_m(P_{X^m})$ independently of $y^n$:
\begin{align}
    \Pr_\omega\left\{x^n(w',\omega) \in \mathcal{C}(y^n)\right\}
    &= \frac{|\mathcal{C}(y^n)|}{|\mathcal{T}_m(P_{X^m})|} \nonumber\\
    &\leq \frac{2^{n(\frac{1}{m}H(P_{X^m})-\hat{I}_m^*(\mathcal{N},P_{X^m}))}
               \cdot (n/m+1)^{\alpha^m\beta^m}}
              {2^{n \cdot \frac{1}{m}H(P_{X^m})} / (n/m+1)^{\alpha^m}} \nonumber\\
    &= 2^{-n\hat{I}_m^*(\mathcal{N},P_{X^m})} \cdot (n/m+1)^{\alpha^m(\beta^m+1)}.
\end{align}
Since $m$ is fixed, $(n/m+1)^{\alpha^m(\beta^m+1)} = 2^{o(n)}$.
A decoding error requires at least one wrong codeword
in $\mathcal{C}(y^n)$, so by the union bound over $M = 2^{nR}-1$ codewords:
\begin{align}
    \Pr_\omega\left\{\text{error} \mid w, v^n\right\}
    &\leq 2^{nR} \cdot 2^{-n\hat{I}_m^*(\mathcal{N},P_{X^m})}
          \cdot (n/m+1)^{\alpha^m(\beta^m+1)} \nonumber\\
    &= 2^{-n(\hat{I}_m^*(\mathcal{N},P_{X^m}) - R)}
       \cdot (n/m+1)^{\alpha^m(\beta^m+1)} \nonumber\\
    &= 2^{-n\eps} \cdot 2^{o(n)} \to 0.
\end{align}
The bound holds for every $v^n \in \mathcal{N}$
(since $\hat{I}_m(x^n;y^n) \geq \hat{I}_m^*(\mathcal{N},P_{X^m})$ for all such $v^n$)
and every message $w$, so $\mathbb{E}_\omega\{\hat{P}_e\} \to 0$. \qedhere
\end{proof}

\begin{remark}[Compatibility of direct and converse.]
\label{rem:gap}
As stated, the direct and converse theorems are proved for
different classes of codes: the converse (Corollary~\ref{cor:capacity-tier1})
applies to deterministic codes, while the achievability (Theorem~\ref{thm:achieve-emp})
uses a randomized code with shared randomness $\omega$ between encoder and decoder,
and establishes small \emph{expected} error
$\mathbb{E}_\omega[\hat{P}_e(w,v^n,\omega)] \to 0$
simultaneously for each fixed $v^n \in \mathcal{N}$ and each message $w$.
Passing from this expected-error guarantee
to a single deterministic codebook that simultaneously achieves
small error for \emph{all} $v^n \in \mathcal{N}$ requires a
union bound over $\mathcal{N}$, which costs $\log|\mathcal{N}| = O(n)$
bits and swamps the $2^{-n\varepsilon}$ gain from the probabilistic
argument; this step remains open for general $\mathcal{N}$.
The same gap is present at the LZ level (Theorems~\ref{thm:channel-tier2}
and~\ref{thm:achievability}).

As mentioned before, it is suggestive to compare this gap to the gap between
direct and converse that exists in the
closely-related individual-sequence channel capacity results of
Lomnitz and Feder~\cite{LF14,LF11} and Shayevitz and Feder~\cite{SF11},
where the direct theorem uses feedback while the converse does not.
We believe that the gap in the present paper is relatively softer for
three reasons: (i)
Common randomness is a less demanding resource than feedback.
Feedback delivers information about past channel outputs to the
encoder at each and every channel use, allowing it to adapt its transmitted
symbols to the realized noise sequence $v^n$; this is a genuine
informational advantage over a deterministic code.
The shared randomness $\omega$, by contrast, is drawn
independently of $v^n$ and conveys no information about the noise;
the encoder cannot exploit $\omega$ to learn or react to $v^n$
in any way. (ii)
Common randomness
can be generated once and for all, and then stored at both encoder
and decoder, after which it is available for all future transmissions
at no additional cost.
Feedback, on the other hand, must be actively communicated from receiver to transmitter
on every channel use, requiring a dedicated noise-free return link and ongoing
overhead proportional to the block length.
(iii) In the classical (probabilistic) channel coding setting
it is known that under mild conditions, common randomness shared between encoder and
decoder does not increase capacity when the noise is chosen
obliviously of the codebook~\cite{BBT60,Ahlswede78,CN88}.
Since in our model $v^n \in \mathcal{N}$ is fixed independently
of $\omega$, the adversary is oblivious to the shared randomness,
and by analogy with these classical results one expects that
$\omega$ does not genuinely increase capacity here either.
Proving this rigorously in the individual-sequence LZ setting
--- i.e., showing that the capacity with shared randomness equals
$C_k(\mathcal{N})$ --- would close the gap and is left as an open
problem.
\eoe
\end{remark}

The coding theorem identifies the
empirical capacity $\hat{C}_m(\calN)$ as the
threshold for reliable communication, but leaves it as an implicit
maximum over input types and minimum over noise sequences.
However, in certain special cases, it is possible to derive a simpler and more
explicit capacity formula. One such special case is the case where $\calN$ is
defined by the set of all $\{v^n\}$ whose empirical entropy does not exceed a
given threshold, say,
\begin{equation}
\label{eq:N-emp-def}
    \calN=\mathcal{N}_{\mathrm{emp}}(H_v) \dfn \{v^n\in\calV^n:~\hHm{m}(v^n) \leq H_v\},
\end{equation}
where $H_v$ is a given constant in the interval $[0,\log\calV]$.
In this case, we next show that
\begin{equation}
\lim_{m\to\infty}\lim_{n\to\infty}\hat{C}_m(\mathcal{N}_{\mathrm{emp}}(H_v))
=\log\alpha-H_v.
\end{equation}

To prove this statement,
the following lemma is the key tool for evaluating this maximum
explicitly for the empirical entropy noise ball: it shows that for a fixed $x^n$,
almost all noise sequences $v^n$ of a given type have empirical conditional entropy
$\hat{H}_m(v^n|x^n)$ close to the unconditional type entropy
$H(P_{V^m})$, i.e., conditioning on $x^n$ essentially does not reduce the
empirical entropy.
This decoupling of $v^n$ from $x^n$ at the empirical level is what
allows the explicit formula $\hat{C}_m(\mathcal{N}_{\mathrm{emp}}(H_v))
= \log\alpha - H_v$ to be derived in
Remark~\ref{rem:closed-form} below.

\begin{lemma}[Empirical Independence]
\label{lem:emp-indep}
Let $\calX$, $\calV$ be finite alphabets. Fix any $x^n \in \calX^n$
and any $m$-block type $P_{V^m}$ on $\calV^m$.
Let $N = n/m$ (assumed integer). For any $\varepsilon > 0$:
\begin{equation}
\label{eq:emp-indep}
    \frac{\#\bigl\{v^n \in \calT(P_{V^m}) :
    \hat{H}_m(v^n \mid x^n) < H(P_{V^m}) - \varepsilon\bigr\}}
    {|\calT(P_{V^m})|}
    \;\leq\;
    (N+1)^{|\calX|^m|\calV|^m+|\calV|^m}
    \cdot 2^{-N\varepsilon},
\end{equation}
which vanishes exponentially in $N$.
Moreover, $\hat{H}_m(v^n \mid x^n) \leq H(P_{V^m})$ holds
for \emph{every} $v^n \in \calT(P_{V^m})$ without exception.
\end{lemma}

\begin{proof}
Treat $m$-blocks as super-letters; there are $N = n/m$ super-letter
positions.

\emph{Upper bound (exact, no exceptions).}
Since conditioning reduces empirical entropy,
$\hat{H}_m(v^n \mid x^n) \leq \hat{H}_m(v^n) = H(P_{V^m})$
for every $v^n \in \calT(P_{V^m})$.

\emph{Lower bound (for most $v^n$).}
By the method of types~\cite{Csiszar98}, the number of $v^n \in
\calT(P_{V^m})$ sharing a given joint type $\hat{P}_{X^m V^m}$
(with $V^m$-marginal $P_{V^m}$) is at most
$2^{N H(\hat{Q}|\hat{P}_{X^m})} = 2^{N\hat{H}_m(v^n|x^n)}$,
and the number of distinct joint types is at most
$(N+1)^{|\calX|^m|\calV|^m}$.
Hence the number of $v^n \in \calT(P_{V^m})$ with
$\hat{H}_m(v^n|x^n) < H(P_{V^m}) - \varepsilon$ is at most
$(N+1)^{|\calX|^m|\calV|^m} \cdot 2^{N(H(P_{V^m})-\varepsilon)}$.
Since $|\calT(P_{V^m})| \geq 2^{NH(P_{V^m})}/(N+1)^{|\calV|^m}$
by the method of types, taking the ratio gives~\eqref{eq:emp-indep}.
\end{proof}

\begin{remark}[Explicit capacity formula]
\label{rem:closed-form}
Under symbol-wise invertibility (Definition~\ref{def:sym-inv}) and
with no input constraints:
\begin{equation}
\label{eq:cap-emp-formula}
    \lim_{m\to\infty}\lim_{n\to\infty}\hat{C}_m\!\left(\mathcal{N}_{\mathrm{emp}}(H_v)\right)
    = \log\alpha - H_v,
\end{equation}
achieved by any $x^n$ with uniform empirical distribution
($\hHm{m}(x^n) = \log\alpha$).
This holds for all symbol-wise invertible channels,
including group-additive channels $y = x \oplus v$.
\end{remark}

\begin{proof}[Proof of Remark~\ref{rem:closed-form}]
By the relabeling lemma~\eqref{eq:minimax-decomp},
$\hat{I}_m(x^n;y^n) = \hHm{m}(y^n) - \hHm{m}(v^n \mid x^n)$
for every $x^n, v^n$.

\emph{Lower bound (exact).}
Choose $x^n$ uniform over $\calX^m$-blocks, so $\hHm{m}(x^n) = \log\alpha$.
Symbol-wise invertibility and uniformity of $x^n$ imply $\hHm{m}(y^n) = \log\alpha$.
For every $v^n \in \mathcal{N}_{\mathrm{emp}}(H_v)$,
$\hHm{m}(v^n \mid x^n) \leq \hHm{m}(v^n) \leq H_v$, so
\[
    \hat{I}_m(x^n;y^n)
    = \hHm{m}(y^n) - \hHm{m}(v^n \mid x^n)
    \geq \log\alpha - H_v,
\]
and therefore $\hat{C}_m(\mathcal{N}_{\mathrm{emp}}(H_v)) \geq \log\alpha - H_v$.

\emph{Upper bound (asymptotic).}
For any $x^n$ and any type $P_{V^m}$ with $H(P_{V^m}) = H_v$,
Lemma~\ref{lem:emp-indep} gives $\hHm{m}(v^n \mid x^n) \geq H_v - \varepsilon$
for all but exponentially few $v^n \in \calT(P_{V^m})$.
Any such $v^n$ lies in $\mathcal{N}_{\mathrm{emp}}(H_v)$ (since $\hHm{m}(v^n) = H_v$), and
\[
    \hat{I}_m(x^n;y^n)
    = \hHm{m}(y^n) - \hHm{m}(v^n \mid x^n)
    \leq \log\alpha - H_v + \varepsilon.
\]
Since $x^n$ is arbitrary and $\varepsilon \to 0$ as $n\to\infty$ then $m\to\infty$,
$\lim_{m\to\infty}\lim_{n\to\infty}\hat{C}_m(\mathcal{N}_{\mathrm{emp}}(H_v)) \leq \log\alpha - H_v$,
which together with the lower bound gives~\eqref{eq:cap-emp-formula}.
\end{proof}

The expression $\log\alpha - H_v$ essentially agrees with the capacity
formula of Shayevitz and Feder~\cite{SF11} and Lomnitz and
Feder~\cite{LF14,LF11} for group-additive channels. There is a slight difference that in their case,
the limiting empirical entropy of the noise, $H_v$, is replaced by the LZ complexity of the noise, but
the two measures are essentially equivalent for large $m$ as will be discussed in the sequel (see also \cite{p246}).

\textbf{Input constraints.}
When $x^n$ is restricted to a set $\mathcal{X}_{\mathrm{cst}}$
(e.g., a power constraint or a composition constraint),
the capacity becomes
\begin{equation}
\label{eq:cap-emp-constrained}
    \hat{C}_m\!\left(\mathcal{N}_{\mathrm{emp}}(H_v),
    \mathcal{X}_{\mathrm{cst}}\right)
    \;=\;
    \max_{x^n \in \mathcal{X}_{\mathrm{cst}}}
    \min_{v^n \in \mathcal{N}_{\mathrm{emp}}(H_v)}
    \hat{I}_m(x^n;y^n).
\end{equation}
By~\eqref{eq:minimax-decomp} this equals
\begin{equation}
    \max_{x^n \in \mathcal{X}_{\mathrm{cst}}}
    \min_{v^n \in \mathcal{N}_{\mathrm{emp}}(H_v)}
    \Bigl[\hHm{m}(y^n) - \hHm{m}(v^n|x^n)\Bigr],
\end{equation}
where $y_t = f(x_t,v_t)$ symbol-wise.
Both terms depend on the noise $v^n$: the output entropy
$\hHm{m}(y^n)$ depends on how $v^n$ interacts with $x^n$
through the channel, while $\hHm{m}(v^n|x^n)$ measures
the residual noise entropy given the input.
The adversary minimizes the difference by choosing $v^n$
that simultaneously reduces the output entropy and
concentrates the noise given the input.
This is the individual-sequence analogue of the classical formula
$C = \max_{P_X} I(X;Y) = \max_{P_X}[H(Y) - H(Y|X)]$,
where the output entropy $H(Y)$ is jointly determined by
the input distribution and the noise statistics.
In particular, an input constraint that forces $x^n$ away
from uniform will in general reduce $\hHm{m}(y^n)$ below
$\log\alpha$, lowering the capacity below $\log\alpha - H_v$.

\section{LZ Channel Capacity}
\label{sec:lz-capacity}

The coding theorem of the preceding section is formulated in terms of
the empirical entropies of $m$-blocks, where $m$ is the block length of the
converse's block-FSM decoder and $s$ is its state count.
These parameters determine the class of reference decoders: larger $m$
and $s$ yield a richer class and hence a tighter converse bound.
However, $m$ and $s$ are not intrinsic parameters of the channel or the noise set
$\mathcal{N}$, and it is desirable to pass to a characterization that is free
of them without taking the explicit limits of $m$ and $s$ tending to infinity.
In this context, prior works of Shayevitz and Feder~\cite{SF11}
and Lomnitz and Feder~\cite{LF14,LF11,LF13b} essentially suggest a
language that is free of those parameters as
their results are stated directly in terms of LZ
complexities. This suggests that we attempt to
reformulate our results in this LZ language as well. This will also
allow a direct
comparison with those works.

It turns out that the replacement of empirical entropies by LZ complexities
require a considerable level of caution. The correct passage to LZ turns out
to be selective. We were not able to replace both the conditional and the
unconditional entropies by the respective LZ and conditional LZ complexities,
but the resulting
capacity $C_k(\mathcal{N})$ turns out to admit a \emph{hybrid} form:
essentially only the conditional entropy of the empirical mutual information can be replaced by conditional LZ
complexity, whereas the unconditional entropy remains essentially intact
(except the block length). 
This hybrid form --- input entropy minus conditional LZ complexity ---
reflects exactly what
the type-class random coding argument requires.
In the further limit $k\to\infty$, both $H(P_{X^k})/k$ and
$\bar{\rho}_k(x^n|y^n)$ converge
to the same LZ complexities, giving the parameter-free expression
$C_\LZ(\mathcal{N})$.

\subsection{LZ Complexity Background}
\label{sec:lz-background}

\subsubsection{LZ78 Complexity}

The \emph{LZ78 incremental parsing}~\cite{ZL78} of $x^n$
proceeds greedily: at each step, find the shortest
extension $p_j$ of a previously seen phrase
$p_j' \in \{p_1,\ldots,p_{j-1}\}$ by one new symbol,
such that $p_j \notin \{p_1,\ldots,p_{j-1}\}$.
Let $c(x^n)$ denote the number of phrases.
The \emph{LZ78 complexity} is:
\begin{equation}
\label{eq:lz-def}
    \rho_\LZ(x^n) \dfn \frac{c(x^n)\log c(x^n)}{n}.
\end{equation}
LZ78 is universally optimal over all finite-state compressors~\cite{ZL78}:
$\rho_\LZ(x^n) \leq \rho_s(x^n) + \eps(n,s)$ where $\eps(n,s)\to 0$ for
fixed $s$, and $\rho_s(x^n)$ is the $s$-state compressibility of $x^n$.

\subsubsection{Conditional LZ Complexity}

Given side information $y^n$, the \emph{conditional LZ
complexity} $\rho_\LZ(x^n|y^n)$ is defined via the
conditional incremental parsing of~\cite{Ziv85}.
Concretely, apply the LZ78 incremental parsing algorithm to the
sequence of pairs $(x_1,y_1),\ldots,(x_n,y_n)$ over
$\calX\times\calY$, obtaining $c(x^n,y^n)$ joint phrases.
Denote by $c'(y^n)$ the number of phrases in the LZ78 parsing
of $y^n$ alone, and by $c_l(x^n|y^n)$ the number of
$x$-components (sub-phrases) within the $l$-th $y$-phrase.
Then:
\begin{equation}
\label{eq:lz-cond-def}
    \rho_\LZ(x^n|y^n) \dfn
    \frac{1}{n}\sum_{l=1}^{c'(y^n)}
    c_l(x^n|y^n)\log c_l(x^n|y^n).
\end{equation}
For fixed $y^n$, $x^n \mapsto \rho_\LZ(x^n|y^n)$ is
approximately the normalized length of a UD code
(up to a lower-order redundancy term~\cite{Ziv85}),
a property used in the achievability proof.

\subsubsection{Block-Average LZ Complexity}

For a given positive integer $k$ dividing $n$ (the LZ-scale block length), we
define the following quantities:
\begin{align}
\label{eq:bar-rho}
    \bar{\rho}_k(x^n) &\dfn
    \frac{k}{n}\sum_{i=0}^{n/k-1}
    \rho_\LZ(x_{ik+1}^{ik+k}),\\
\label{eq:bar-rho-cond}
    \bar{\rho}_k(x^n|y^n) &\dfn
    \frac{k}{n}\sum_{i=0}^{n/k-1}
    \rho_\LZ(x_{ik+1}^{ik+k}|y_{ik+1}^{ik+k}),\\
\label{eq:bar-I}
    \bar{I}_k(x^n;y^n) &\dfn
    \bar{\rho}_k(x^n) - \bar{\rho}_k(x^n|y^n).
\end{align}
The \emph{$k$-block resetting scheme} refers throughout the sequel to the
application of the LZ78 algorithm independently in successive blocks, with
the LZ dictionary reset at the start of each $k$-block.

\subsubsection{Sandwiching Bounds}
\label{sec:sandwich}

The following theorem gives the upper sandwiching bound
that connects $\bar{\rho}_k$ to empirical entropy,
used in the converse proof below.

\begin{theorem}[Upper sandwiching bound~\cite{p246}]
\label{thm:sandwich}
For every $x^n \in \calX^n$, positive integer $k$ dividing $n$, and $m \geq 1$:
\begin{equation}
    \bar{\rho}_k(x^n)
    \leq \hHm{m}(x^n) + \frac{1}{m}
    + \frac{2(m+1)\alpha^{m+1}}{n}\log\frac{n}{m}
    + \eps_2(k,\alpha^{2m}),
    \label{eq:sand-upper}
\end{equation}
where $\eps_2(k,s)\to 0$ as $k\to\infty$ for fixed $s$.
An analogous bound holds for conditional quantities,
with $\alpha\beta$ replacing $\alpha$ in the exponents.
\end{theorem}

We abbreviate~\eqref{eq:sand-upper} as:
\begin{equation}
\label{eq:Dm}
    \bar{\rho}_k(x^n) \leq \hHm{m}(x^n)
    + \delta_+(k,m,n,\alpha),
    \quad
    \bar{\rho}_k(x^n|y^n) \leq \hHm{m}(x^n|y^n)
    + \delta_+(k,m,n,\alpha\beta),
\end{equation}
where $\delta_+(k,m,n,\cdot)\to 0$ as $n \gg k \gg m \gg 1$ with $\log n \gg m$.
Here `$k \gg m$' means $k/m \to \infty$, which is needed for
$\eps_2(k,\alpha^{2m}) \to 0$: since $\eps_2(k,s) = O((\log s)/k)$
(by the LZ redundancy bound~\cite{ZL78}), we have
$\eps_2(k,\alpha^{2m}) = O(m\log\alpha/k) \to 0$ whenever $k/m \to \infty$.
The condition $\log n \gg m$ (equivalently $m = o(\log n)$) is
needed for the term $2(m+1)\alpha^{m+1}/n \cdot \log(n/m) \to 0$.

The conditional bound~\eqref{eq:Dm} is the sole bridge used in the
converse proof (Step~4 of Theorem~\ref{thm:channel-tier2} below):
it converts the Fano-based bound on $\hHm{m}(x^n|y^n)$ into
a bound on $\bar{\rho}_k(x^n|y^n)$, under the hierarchy
$n \gg k \gg m \gg 1$, $m = o(\log n)$.
No sandwiching is needed for the achievability, which operates
directly at the LZ level.

\subsection{LZ Capacity: Definition and Converse}
\label{sec:lz-converse}

\begin{definition}[Individual-sequence LZ channel capacity]
\label{def:capacity-k}
Let $\mathcal{N} \subseteq \calV^n$ be any noise set.
For a channel (arbitrary deterministic mapping
$x^n \mapsto y^n$, parameterized by $v^n \in \mathcal{N}$)
and LZ-scale block length $k$ (with $k | n$), the \emph{block-average LZ channel
capacity} is:
\begin{equation}
\label{eq:capacity-k}
    C_k(\mathcal{N}) \dfn
    \sup_{x^n \in \calX^n}
    \min_{v^n \in \mathcal{N}}
    \left[\frac{H(P_{X^k})}{k} - \bar{\rho}_k(x^n|y^n)\right],
\end{equation}
where $P_{X^k}$ is the $k$-block empirical distribution of $x^n$,
and $y^n = q(x^n,v^n)$ is the channel output~\eqref{eq:general-channel}.
This is well-defined for any $\mathcal{N} \subseteq \calV^n$: the
inner expression is a real number for each pair $(x^n, v^n)$, and the
minimum over the finite set $\mathcal{N}$ and supremum over the finite
set $\calX^n$ are unambiguous.
\end{definition}

This is the individual-sequence analogue of the compound
channel capacity: maximize over input distributions (here, $k$-block
types), minimize over noise sequences in $\mathcal{N}$.
No cardinality condition on $\mathcal{N}$ is needed
for the converse to hold.
The choice of $\mathcal{N}$ is flexible; the converse
(Theorem~\ref{thm:channel-tier2}) applies to \emph{any}
$\mathcal{N} \subseteq \calV^n$.
The achievability (Theorem~\ref{thm:achievability})
additionally requires $\mathcal{N}$ to be closed under $k$-block
permutations: for any $v^n \in \mathcal{N}$ and any
$k$-block permutation $\pi$, the permuted sequence
$v^n_\pi$ must also belong to $\mathcal{N}$.
This condition is satisfied by $\mathcal{N}_{\mathrm{LZ}}(\rho_v)$
(since $\bar{\rho}_k$ is permutation-invariant), by Hamming balls,
and by any set defined by a permutation-invariant criterion.
The classical LZ ball $\{v^n:\rho_\LZ(v^n)\leq \rho_v\}$ is
\emph{not} permutation-closed in general (a permutation may increase
$\rho_\LZ$ by disrupting the global parsing structure), so
achievability for that set remains an open problem.

\begin{theorem}[Channel capacity converse in LZ terms]
\label{thm:channel-tier2}
Let the channel be $y^n = q(x^n,v^n)$ for an arbitrary
deterministic function $q$, with $v^n \in \mathcal{N}$
for some noise set $\mathcal{N} \subseteq \calV^n$.
Let the encoder be arbitrary (no finite-state constraint)
and the decoder be a block-FSM with $s$ states and block length $m$.
Suppose a codebook of $M = 2^{nR}$ codewords achieves small
$\hat{P}_e$ (average symbol-error frequency, as in~\eqref{eq:pe-def}) for every
$v^n \in \mathcal{N}$.
Then for positive integers $k \gg m \gg 1$
(with $m | n$, $k | n$, and $\log n \gg m$):
\begin{equation}
\label{eq:channel-tier2}
    \phi(\hat{P}_e) \geq
    \bar{\rho}_k(x^n|y^n) - \Gamma_0,
\end{equation}
and consequently:
\begin{equation}
\label{eq:capacity-bound-tier2}
    R \leq C_k(\mathcal{N})
    + \phi\!\left(\max_{v^n \in \mathcal{N}}\hat{P}_e(v^n)\right) + \Gamma_0,
\end{equation}
where $C_k(\mathcal{N})$ is defined in Definition~\ref{def:capacity-k}, and
\begin{equation}
\label{eq:Gamma0-def}
    \Gamma_0 = \frac{\log s}{m} + \delta_+(k,m,n,\alpha\beta),
\end{equation}
with $\Gamma_0 \to 0$ as $k \gg m \gg 1$ with $\log n \gg m$
and $s$, $\alpha$, $\beta$ fixed.
\end{theorem}

\begin{proof}
\textbf{Step 1: Dominant type and rate bound.}
By the same pigeonhole argument as in the proof of
Corollary~\ref{cor:capacity-tier1} (with $m$ replaced by $k$
and order-$m$ types replaced by $k$-block types), there exists
a $k$-block type $P_{X^k}$ such that at least $2^{nR-o(n)}$
codewords share type $P_{X^k}$, and
\begin{equation}
\label{eq:ct2-rate-bound}
    R \leq \frac{H(P_{X^k})}{k} + o(1).
\end{equation}
Fix this dominant type $P_{X^k}$ and any codeword
$x^n$ of this type for the remainder of the proof.

\textbf{Step 2: Worst noise and Fano.}
Let $v_n^*$ be a maximizer of $\bar{\rho}_k(x^n|q(x^n,v^n))$
across $\calN$ for the fixed codeword $x^n$ of Step~1.
Since $\phi$ is non-decreasing,
\begin{equation}
\label{eq:ct2-worst}
    \phi\!\left(\max_{v^n \in \mathcal{N}} \hat{P}_e(v^n)\right)
    \geq \phi(\hat{P}_e(v^n_*))
\end{equation}
so it suffices to lower-bound $\phi(\hat{P}_e(v^n_*))$.
In what follows, $y^n = q(x^n, v^n_*)$.

\textbf{Step 3: Apply Theorem~\ref{thm:main-tier2}.}
Theorem~\ref{thm:main-tier2} (with block length $m$,
decoder state count $s$) gives:
\begin{equation}
\label{eq:ct2-step1}
    \phi(\hat{P}_e(v^n_*)) \geq \hHm{m}(x^n|y^n) - \frac{\log s}{m}.
\end{equation}

\textbf{Step 4: Convert empirical entropy to $\bar{\rho}_k$ via sandwiching.}
By the upper sandwiching bound~\eqref{eq:Dm} applied to $(x^n \mid y^n)$
(joint alphabet size $\alpha\beta$):
\begin{equation}
\label{eq:ct2-step2}
    \bar{\rho}_k(x^n|y^n) \leq \hHm{m}(x^n|y^n) + \delta_+(k,m,n,\alpha\beta),
\end{equation}
so $\hHm{m}(x^n|y^n) \geq \bar{\rho}_k(x^n|y^n) - \delta_+(k,m,n,\alpha\beta)$.
Substituting into~\eqref{eq:ct2-step1}:
\begin{equation}
\label{eq:ct2-combined}
    \phi(\hat{P}_e(v^n_*)) \geq \bar{\rho}_k(x^n|y^n) - \frac{\log s}{m} - \delta_+(k,m,n,\alpha\beta)
    = \bar{\rho}_k(x^n|y^n) - \Gamma_0,
\end{equation}
where $\Gamma_0 = (\log s)/m + \delta_+(k,m,n,\alpha\beta) \to 0$ as $k \gg m \gg 1$
with $\log n \gg m$ and $s$ fixed.
Combined with~\eqref{eq:ct2-worst}, this gives~\eqref{eq:channel-tier2}.

\textbf{Step 5: Upper-bound $R$ using $C_k(\mathcal{N})$.}
Combining~\eqref{eq:ct2-rate-bound} and~\eqref{eq:ct2-combined}
with~\eqref{eq:ct2-worst}:
\begin{align}
    R &\leq \frac{H(P_{X^k})}{k} + o(1) \notag \\
      &= \frac{H(P_{X^k})}{k} - \bar{\rho}_k(x^n|y^n)
         + \bar{\rho}_k(x^n|y^n) + o(1) \notag \\
      &\leq \frac{H(P_{X^k})}{k} - \bar{\rho}_k(x^n|y^n)
         + \phi\!\left(\max_{v^n\in\calN}\hat{P}_e(v^n)\right)
         + \Gamma_0 + o(1).
\end{align}
By definition of $C_k(\mathcal{N})$
(Definition~\ref{def:capacity-k}),
\begin{eqnarray}
    \frac{H(P_{X^k})}{k} - \bar{\rho}_k(x^n|y^n)
    &=&\frac{H(P_{X^k})}{k} -
\max_{v^n\in\mathcal{N}}\bar{\rho}_k(x^n|q(x^n,v^n))\nonumber\\
    &=&\min_{v^n\in\mathcal{N}}\left[\frac{H(P_{X^k})}{k} -
\bar{\rho}_k(x^n|q(x^n,v^n))\right]\nonumber\\
    &\leq&\max_{\tilde{x}^n \in \calX^n}\min_{v^n\in\mathcal{N}}
    \left[\frac{H(P_{\tilde{X}^k})}{k} -
\bar{\rho}_k(\tilde{x}^n|q(\tilde{x}^n,v^n))\right]\nonumber\\
    &=&C_k(\mathcal{N}).
\end{eqnarray}
Absorbing the $o(1)$ into $\Gamma_0$ gives~\eqref{eq:capacity-bound-tier2}.
\end{proof}

\begin{remark}[Operational interpretation via predictive coding]
\label{rem:predictive-coding}
Theorem~\ref{thm:channel-tier2} admits a direct operational
interpretation in terms of \emph{predictive coding with a
finite-state predictor}, offering a perspective complementary
to the sandwich-based proof above.

Consider any two-stage encoder for $x^n$ with side information $y^n$:
\begin{enumerate}
\item \emph{Prediction:} Apply a block-FSM filter with $s$ states
to $y^n$ to produce a prediction $\hat{x}^n$ of $x^n$.
\item \emph{Residual coding:} Compress the error sequence
$e^n = x^n - \hat{x}^n$ (well-defined w.l.o.g., since Hamming
distance is a difference distortion measure) using a memoryless
empirical entropy coder at rate $\hat{H}_1(e^n)$,
plus a redundancy term that vanishes with $n$.
\end{enumerate}
Since $e_t = 0$ whenever $\hat{x}_t = x_t$, the symbol $e_t$
takes value $0$ with empirical frequency $1 - \hat{P}_e$.
The empirical entropy $\hat{H}_1(e^n)$ is therefore at most
$\phi(\hat{P}_e) = h_2(\hat{P}_e) + \hat{P}_e\log(\alpha-1)$,
the maximum entropy of any distribution on $\calX$ with
$P(e=0) = 1-\hat{P}_e$.

On the other hand, by~\cite{me00}, no finite-state encoder
with side information $y^n$ can compress $x^n$ to a rate
below $\bar{\rho}_k(x^n|y^n)$ minus a redundancy term
(arising from finite-$k$ and finite-$n$ effects, of the
same character as $\delta_+$ in~\eqref{eq:Dm}).
Since the two-stage encoder is a particular finite-state
encoder, its rate $\hat{H}_1(e^n)$ cannot be below this
fundamental limit.
Absorbing all redundancy terms into a single $\Gamma_0$
(of the same character as in~\eqref{eq:Gamma0-def}) then
yields a bound of the same form as~\eqref{eq:channel-tier2},
though the precise redundancy term may differ.

Thus, Theorem~\ref{thm:channel-tier2} is the statement that
\emph{no finite-state predictor followed by memoryless
residual coding can achieve a rate below $\phi(\hat{P}_e)$}
--- regardless of the predictor used.
\end{remark}

\subsection{Achievability}
\label{sec:achievability}

We prove achievability for the channel model
$y^n = q(x^n,v^n)$ when the encoder knows $q$ and
$\mathcal{N}$ but not the specific $v^n \in \mathcal{N}$.
Throughout this subsection the noise set $\mathcal{N}$ is
assumed to be \emph{closed under $k$-block permutations}:
for any $v^n \in \mathcal{N}$ and any permutation $\pi$ of
the $n/k$ consecutive $k$-blocks of $v^n$, we have
$v^n_\pi \in \mathcal{N}$.
This is the structural condition needed for the random-coding
argument; it holds for all noise sets listed in the
examples following Definition~\ref{def:capacity-k}.
The main result uses capacity $C_k(\mathcal{N})$
(Definition~\ref{def:capacity-k}).

\begin{definition}[$k$-block resetting conditional LZ decoder]
\label{def:decoder}
Fix a positive integer $k$ dividing $n$.
Given a codebook $\mathcal{C}_n \subset \calX^n$ and channel
output $y^n$, the \emph{$k$-block resetting conditional LZ
decoder} computes, for each $x^n \in \mathcal{C}_n$, the
block-average conditional LZ complexity $\bar{\rho}_k(x^n|y^n)$
as defined in~\eqref{eq:bar-rho-cond}, where each block's
conditional LZ complexity is computed independently (resetting
to a fresh empty dictionary at each block boundary), and outputs:
\begin{equation}
\label{eq:decoder}
    \hat{x}^n \dfn \arg\min_{x^n \in \mathcal{C}_n}
    \bar{\rho}_k(x^n \mid y^n),
\end{equation}
with ties broken arbitrarily.
\eoe
\end{definition}

\begin{remark}[Type-independence of the inner minimum for permutation-closed $\mathcal{N}$]
\label{rem:capacity-well-defined}
Since $\mathcal{N}$ is closed under $k$-block permutations (as assumed
throughout this subsection), the inner minimum in~\eqref{eq:capacity-k}
depends only on the $k$-block type of $x^n$, not on the specific
representative.
Indeed, if $x^n$ and $\tilde{x}^n$ share the same $k$-block type
$P_{X^k}$, then $\tilde{x}^n = x^n_\pi$ for some $k$-block
permutation $\pi$, and the map $v^n \mapsto v^n_\pi$ is a bijection on
$\mathcal{N}$, so
\[
    \min_{v^n \in \mathcal{N}} \bar{\rho}_k(\tilde{x}^n \mid q(\tilde{x}^n, v^n))
    = \min_{v^n \in \mathcal{N}} \bar{\rho}_k(x^n_\pi \mid q(x^n_\pi, v^n_\pi)).
\]
Since $\bar{\rho}_k(\cdot|\cdot)$ is a block average, it is invariant
under joint $k$-block permutation of both arguments:
$\bar{\rho}_k(x^n_\pi \mid q(x^n_\pi, v^n_\pi))
= \bar{\rho}_k(x^n \mid q(x^n, v^n))$.
Hence the minimum over $\mathcal{N}$ is the same for all representatives
of a given $k$-block type, and the supremum in~\eqref{eq:capacity-k}
reduces equivalently to a supremum over $k$-block types.
\end{remark}

\begin{definition}[Randomized codebook ensemble]
\label{def:ensemble}
Fix a channel $q$, noise set $\mathcal{N}$,
LZ-scale block length $k$, and rate $R < C_k(\mathcal{N})$.
Since the number of $k$-block types is finite, the supremum
in~\eqref{eq:capacity-k} is attained; let $P^*_{X^k}$ be any
maximizing type:
\[
    \min_{v^n\in\mathcal{N}}
    \left[\frac{H(P^*_{X^k})}{k} - \bar{\rho}_k(x^n|q(x^n,v^n))\right]
    = C_k(\mathcal{N}),
\]
for any $x^n$ with $k$-block empirical distribution $P^*_{X^k}$
(by Remark~\ref{rem:capacity-well-defined},
the value does not depend on the choice of representative).
The \emph{$k$-type class} is
\[
    \calT_k(P^*_{X^k}) \dfn
    \bigl\{x^n \in \calX^n : P_{X^k}(x^n) = P^*_{X^k}\bigr\},
\]
i.e., all sequences whose $k$-block empirical distribution equals $P^*_{X^k}$.
Set $M \dfn \lfloor 2^{nR}\rfloor$.
A \emph{randomized code} consists of:
\begin{itemize}
    \item A \emph{message set}
    $\mathcal{M}_n \dfn \{1, 2, \ldots, M\}$.
    \item An \emph{ensemble of codebooks}
    $\{\mathcal{C}_n(\omega)\}_\omega$, each consisting of
    $M$ codewords drawn independently and uniformly from
    $\calT_k(P^*_{X^k})$, indexed by a shared random key
    $\omega$ drawn uniformly from a finite index set.
    \item A \emph{decoder}: the $k$-block resetting
    conditional LZ decoder
    (Definition~\ref{def:decoder}), applied to the
    codebook $\mathcal{C}_n(\omega)$.
\end{itemize}
The key $\omega$ is known to both encoder and decoder.
\eoe
\end{definition}


\begin{lemma}[Kraft counting lemma]
\label{lem:kraft}
For any fixed $y^n \in \calY^n$, positive integer $k$
dividing $n$, and any $L \geq 0$:
\begin{equation}
\label{eq:kraft-count}
    \bigl|\bigl\{\tilde{x}^n \in \calX^n :
    \bar{\rho}_k(\tilde{x}^n|y^n) \leq \frac{L}{n}\bigr\}\bigr|
    \leq 2^{L + o(n)},
\end{equation}
where $o(n)$ depends only on $n$, $k$, and $|\calX|$
(not on $y^n$ or $\tilde{x}^n$).
\end{lemma}

\begin{proof}
The $k$-block resetting conditional LZ scheme
(Definition~\ref{def:decoder}) encodes $\tilde{x}^n$ given $y^n$
block by block, each block independently.
For block $i$ (indices $ik+1$ to $(i+1)k$), the
conditional LZ algorithm~\cite{Ziv85} on $k$ symbols
produces a uniquely decodable code for $\tilde{x}_{ik+1}^{(i+1)k}$
given ${y^n}_{ik+1}^{(i+1)k}$ with codelength
$\ell_i(\tilde{x}^n) \dfn k\rho_\LZ(\tilde{x}_{ik+1}^{(i+1)k} \mid {y^n}_{ik+1}^{(i+1)k})
+ O(c_i)$,
where $c_i \leq k/\log k$ is the number of phrases and
$O(c_i) = o(k)$.
Concatenating the $n/k$ independent block codes gives a
UD code for $\tilde{x}^n$ given $y^n$ with total codelength
$\ell(\tilde{x}^n) = \sum_{i=0}^{n/k-1}\ell_i(\tilde{x}^n)
= n\bar{\rho}_k(\tilde{x}^n|y^n) + o(n)$.
By the Kraft inequality applied independently within each
block (the blocks are decoded independently):
\begin{equation}
    \sum_{\tilde{x}^n \in \calX^n} 2^{-\ell(\tilde{x}^n)} \leq 1.
\end{equation}
For $\tilde{x}^n$ with $\bar{\rho}_k(\tilde{x}^n|y^n) \leq L/n$:
$\ell(\tilde{x}^n) \leq L + o(n)$.
Restricting the Kraft sum:
\begin{equation}
    \bigl|\bigl\{\tilde{x}^n : \bar{\rho}_k(\tilde{x}^n|y^n) \leq
    \tfrac{L}{n}\bigr\}\bigr| \cdot 2^{-(L+o(n))}
    \leq \sum_{\tilde{x}^n:\bar{\rho}_k(\tilde{x}^n|y^n)\leq L/n}
    2^{-\ell(\tilde{x}^n)} \leq 1,
\end{equation}
giving~\eqref{eq:kraft-count}.
\end{proof}

\begin{theorem}[Achievability]
\label{thm:achievability}
Let the channel be symbol-wise invertible
(Definition~\ref{def:sym-inv}) and let $\mathcal{N}$ be a
noise set closed under $k$-block permutations:
for every $v^n \in \mathcal{N}$ and every permutation $\pi$ of the
$n/k$ consecutive $k$-blocks of $v^n$, the permuted sequence
$v^n_\pi$ belongs to $\mathcal{N}$.
The capacity $C_k(\mathcal{N})$ is defined in Definition~\ref{def:capacity-k}.
For any rate $R < C_k(\mathcal{N})$ and all
sufficiently large $n$, there exists a randomized code
of $2^{nR}$ messages and block length $n$
(with the $k$-block resetting conditional LZ
decoder of Definition~\ref{def:decoder}) such that the
expected per-symbol error frequency satisfies
\begin{equation}
\label{eq:exp-Pe}
    \max_{i \in \mathcal{M}_n}\mathbb{E}_\omega\bigl[\hat{P}_e(i)\bigr]
    \;\to\; 0
\end{equation}
simultaneously for every $v^n \in \mathcal{N}$,
where $\eps \dfn C_k(\mathcal{N}) - R > 0$.
\end{theorem}

\begin{proof}[Proof of Theorem~\ref{thm:achievability}]
Let $\eps \dfn C_k(\calN) - R > 0$.

\medskip
\noindent\textbf{Step 1: Capacity-achieving type and $k$-type class.}
Since the number of $k$-block types is finite, the supremum in $C_k(\calN)$
is attained; let $P^*_{X^k}$ be a maximizing type
(Definition~\ref{def:ensemble}), so that for any $x^n \in \calT_k(P^*_{X^k})$:
\begin{equation}
\label{eq:ach-cond-ub}
    \bar{\rho}_k(x^n|\, q(x^n,v^n))
    \leq \frac{H(P^*_{X^k})}{k} - C_k(\calN)
    \qquad \forall\, v^n \in \calN,
\end{equation}
where we used Remark~\ref{rem:capacity-well-defined}: since $\calN$ is
closed under $k$-block permutations, the inner minimum in $C_k(\calN)$
is the same for all sequences in $\calT_k(P^*_{X^k})$.
The $k$-type class $\calT_k(P^*_{X^k})$ is defined in
Definition~\ref{def:ensemble}.
By the method of types applied to the alphabet $\calX^k$ with block length $n/k$,
\begin{equation}
\label{eq:type-lb}
    |\calT_k(P^*_{X^k})| \;\geq\; 2^{(n/k) H(P^*_{X^k}) - o(n/k)}
    \;=\; 2^{n H(P^*_{X^k})/k - o(n)}.
\end{equation}
Since $R < C_k(\calN) \leq H(P^*_{X^k})/k$ (the inequality $C_k(\calN) \leq H(P^*_{X^k})/k$
holds because $\bar{\rho}_k \geq 0$), we have
\begin{equation}
    |\calT_k(P^*_{X^k})| \geq 2^{nR}  = M
\end{equation}
for all sufficiently large $n$, so the ensemble is well-defined.

\medskip
\noindent\textbf{Step 2: Random codebook and decoding rule.}
Draw $M$ codewords $X(1),\ldots,X(M)$
independently and uniformly at random from $\calT_k(P^*_{X^k})$.
The shared randomness $\omega$ is available to encoder and decoder.
On receiving $y^n$, the $k$-block resetting conditional LZ
decoder (Definition~\ref{def:decoder}) outputs
\begin{equation}
\label{eq:ach-decoder}
    \hat{i} \;=\;
    \arg\min_{i \in \calM_n}\; \bar{\rho}_k(X(i) \mid y^n),
\end{equation}
breaking ties arbitrarily.

\medskip
\noindent\textbf{Step 3: Correct codeword score.}
Fix $v^n \in \calN$ and a transmitted codeword $x^n = X(i) \in \calT_k(P^*_{X^k})$,
with $y^n = q(x^n,v^n)$.
We claim:
\begin{equation}
\label{eq:correct-score-lz}
    \bar{\rho}_k(x^n \mid y^n) \leq \frac{H(P^*_{X^k})}{k} - C_k(\calN).
\end{equation}
This follows directly from~\eqref{eq:ach-cond-ub}, which applies to
$x^n$ since $x^n \in \calT_k(P^*_{X^k})$.
Note that the threshold $H(P^*_{X^k})/k - C_k(\calN)$
is the same for \emph{every} $x^n \in \calT_k(P^*_{X^k})$
and every $v^n \in \calN$.

\medskip
\noindent\textbf{Step 4: Confusable set.}
For fixed $y^n$, define the confusable set with threshold from~\eqref{eq:correct-score-lz}:
\begin{equation}
\label{eq:G-def}
    \calG(y^n) \dfn
    \bigl\{\tilde{x}^n \in \calX^n :
    \bar{\rho}_k(\tilde{x}^n \mid y^n)
    \leq \frac{H(P^*_{X^k})}{k} - C_k(\calN)\bigr\}.
\end{equation}
By the Kraft counting Lemma~\ref{lem:kraft}:
\begin{equation}
\label{eq:G-card}
    |\calG(y^n)| \leq 2^{n(H(P^*_{X^k})/k - C_k(\calN)) + o(n)}.
\end{equation}
A wrong codeword $\widetilde{X}^n = X(j)$, $j \neq i$, drawn uniformly
from $\calT_k(P^*_{X^k})$ independently of $y^n$, causes an error
only if $\widetilde{X}^n \in \calG(y^n)$.
Since $\widetilde{X}^n$ is independent of $y^n$:
\begin{equation}
\label{eq:conf-cond}
    \Pr_\omega\bigl[\widetilde{X}^n \in \calG(y^n)\bigm| X^n=x^n\bigr]
    \leq \frac{|\calG(y^n)|}{|\calT_k(P^*_{X^k})|}.
\end{equation}

\medskip
\noindent\textbf{Step 5: Union bound and average error.}
By the union bound over $M-1$ competing codewords,
using~\eqref{eq:G-card}, \eqref{eq:conf-cond},
and $|\calT_k(P^*_{X^k})| \geq 2^{nH(P^*_{X^k})/k+o(n)}$
from~\eqref{eq:type-lb}:
\begin{align}
\label{eq:Pe-final}
    \E_\omega\bigl[\hat{P}_e(i)\bigm| X(i)=x^n\bigr]
    &\leq 2^{nR} \cdot
    \frac{2^{n(H(P^*_{X^k})/k - C_k(\calN)) + o(n)}}
         {2^{nH(P^*_{X^k})/k + o(n)}}
    \nonumber\\
    &= 2^{n(R - C_k(\calN)) + o(n)}
    = 2^{-n\eps + o(n)}.
\end{align}
Hence:
\begin{equation}
\label{eq:Pe-cond}
    \E_\omega\bigl[\hat{P}_e(i)\bigm| X(i)=x^n\bigr]
    \leq 2^{-n\eps + o(n)}
    \qquad \forall\, x^n \in \calT_k(P^*_{X^k}),\; \forall\, v^n \in \calN.
\end{equation}
Since~\eqref{eq:Pe-cond} holds for every $x^n \in \calT_k(P^*_{X^k})$,
taking the expectation over $X(i)$ (which is uniform on $\calT_k(P^*_{X^k})$)
gives $\E_\omega[\hat{P}_e(i)] \leq 2^{-n\eps+o(n)} \to 0$
simultaneously for every message $i \in \calM_n$ and every $v^n \in \calN$,
which is~\eqref{eq:exp-Pe}.
\end{proof}

\begin{remark}[Type-class noise set and universality of the decoder]
\label{rem:type-class-noise}
Consider the case $\mathcal{N} = \mathcal{T}_k(P_{V^k})$, the
$k$-block type class of a given noise type $P_{V^k}$.
This set is closed under $k$-block permutations,
so both the converse
(Theorem~\ref{thm:channel-tier2}) and the achievability
(Theorem~\ref{thm:achievability}) apply.
Moreover, by the same permutation argument as in
Remark~\ref{rem:capacity-well-defined}, $C_k(\mathcal{T}_k(P_{V^k}))$
equals the per-realization capacity $C_k(v^n)$ (i.e., the capacity for the case
$\calN=\{v^n\}$) for every
$v^n \in \mathcal{T}_k(P_{V^k})$: all noise sequences of the same
type yield the same capacity.
This special case brings out two points.

\emph{The converse is non-trivial only because of the block-FSM constraint.}
For a fixed $v^n$, an unconstrained decoder could cancel $v^n$ exactly
(e.g., for a modulo-additive channel, subtract $v^n$ symbol by symbol),
making the effective capacity $\log\alpha$ regardless of $v^n$, and the
converse would be vacuous.
The block-FSM constraint limits the decoder's inter-block memory to
$\log s$ bits, preventing it from storing or globally exploiting the
structure of $v^n$; this is what gives the converse its content even
when $v^n$ is known in principle.

\emph{The achievability is fully universal.}
The $k$-block resetting conditional LZ decoder
(Definition~\ref{def:decoder}) computes $\bar{\rho}_k(\tilde{x}^n \mid y^n)$
from the received $y^n$ alone, with no knowledge of which
$v^n \in \mathcal{T}_k(P_{V^k})$ occurred.
Thus the \emph{decoder} is universal, achieving $C_k(v^n)$ simultaneously
for every noise sequence in the type class.
The \emph{encoder} selects the capacity-achieving type $P^*_{X^k}$, which
depends on $P_{V^k}$ but not on the specific $v^n$; if furthermore the
input type is fixed externally (e.g., by a composition or power constraint),
the encoder too requires no knowledge of $v^n$, and the entire scheme
is universal.

The same observations apply verbatim at the empirical-entropy level
(Theorem~\ref{thm:achieve-emp} and Corollary~\ref{cor:capacity-tier1}),
with $\mathcal{N} = \mathcal{T}_m(P_{V^m})$ and
$\hat{C}_m(\mathcal{T}_m(P_{V^m}))$ in place of $C_k(\mathcal{T}_k(P_{V^k}))$.
\eoe
\end{remark}

\subsection{Application to Joint Source-Channel Coding}
\label{sec:jscc}

Suppose the encoder first compresses a source sequence $u^n \in \mathcal{U}^n$
into a channel input $x^n \in \calX^n$ via a \emph{block finite-state invertible}
(block-FSI) encoder with state count $s$: a causal encoder with $s$ states that
maps $m$-blocks of $u^n$ to $m$-blocks of $x^n$, and whose inverse (recovering
$u^n$ from $x^n$) is also a block-FSM with at most $s$ states.
The resulting $(u^n, x^n, y^n)$ chain realizes joint source-channel coding (JSCC).
The following lemma shows that the empirical entropy is nearly preserved by such
an encoder.

\begin{lemma}[Approximate empirical entropy equivalence]
\label{lem:equiv}
Let $u^n \mapsto x^n$ be a block-FSI encoder with encoder state count $s$
and inverse-decoder state count $s$, and let $m | n$.
Then for every $y^n \in \calY^n$:
\begin{align}
    |\hHm{m}(u^n) - \hHm{m}(x^n)|
    &\leq \frac{2\log s}{m},
    \label{eq:equiv-marg}\\
    |\hHm{m}(u^n|y^n) - \hHm{m}(x^n|y^n)|
    &\leq \frac{2\log s}{m},
    \label{eq:equiv-cond}\\
    |\hat{I}_m(u^n;y^n) - \hat{I}_m(x^n;y^n)|
    &\leq \frac{4\log s}{m}.
    \label{eq:equiv-MI}
\end{align}
All bounds vanish as $m \gg \log s$.
\end{lemma}

\begin{proof}
\textbf{Proof of~\eqref{eq:equiv-marg}.}
Since the encoder state at the start of block $i$ takes at most $s$ values,
$\hHm{m}(x^n|u^n) \leq (\log s)/m$ (given $u^n$, at most $s$ choices for
the encoder state determine $x^n$ block by block).
By the chain rule $\hHm{m}(x^n) \leq \hHm{m}(u^n) + \hHm{m}(x^n|u^n)$,
so $\hHm{m}(x^n) - \hHm{m}(u^n) \leq (\log s)/m$.
By symmetry (the inverse decoder has $s$ states, giving
$\hHm{m}(u^n|x^n) \leq (\log s)/m$):
$\hHm{m}(u^n) - \hHm{m}(x^n) \leq (\log s)/m$.
Combining gives~\eqref{eq:equiv-marg}.

\textbf{Proof of~\eqref{eq:equiv-cond}.}
Since the encoder is block-FSI, the encoder state at the start of each
$m$-block is determined by $u^n$ up to that point.
Hence $\hHm{m}(x^n|u^n, y^n) \leq (\log s)/m$.
Using the chain rule:
\[
    \hHm{m}(x^n|y^n)
    \leq \hHm{m}(u^n|y^n) + \hHm{m}(x^n|u^n,y^n)
    \leq \hHm{m}(u^n|y^n) + \frac{\log s}{m},
\]
and by symmetry $\hHm{m}(u^n|y^n) \leq \hHm{m}(x^n|y^n) + (\log s)/m$.
Combining gives~\eqref{eq:equiv-cond}.

\textbf{Proof of~\eqref{eq:equiv-MI}.}
Follows from~\eqref{eq:equiv-marg} and~\eqref{eq:equiv-cond}
by the triangle inequality applied to
$\hat{I}_m = \hHm{m}(\cdot) - \hHm{m}(\cdot|\cdot)$. \qedhere
\end{proof}

Combining Theorem~\ref{thm:main-tier2} with Lemma~\ref{lem:equiv} yields
an individual-sequence Fano inequality stated directly in terms of the source.

\begin{theorem}[Individual-sequence Fano inequality for JSCC]
\label{thm:channel-tier1-jscc}
Under the conditions of Lemma~\ref{lem:equiv}
(block-FSI encoder $u^n \mapsto x^n$ with state count $s$,
$m | n$),
let $y^n$ be the channel's response to $x^n$.
Let $\hat{u}^n$ be the output of a block-FSM decoder with block length $m$
and state count $s$ applied directly to $y^n$, and define the source
reconstruction error frequency
\[
    \hat{P}_e^u \dfn \frac{1}{n}\sum_{t=1}^n \mathbf{1}\{\hat{u}_t \neq u_t\}.
\]
Then:
\begin{align}
\label{eq:channel-tier1-jscc}
    \phi\!\left(\hat{P}_e^u\right)
    &\geq \hHm{m}(u^n|y^n) - \frac{\log s}{m} \nonumber\\
    &\geq \hHm{m}(u^n) - \hat{I}_m(u^n;y^n) - \frac{\log s}{m}.
\end{align}
In particular, if reliable source reconstruction is required for every
$v^n \in \mathcal{N}$:
\begin{align}
\label{eq:jscc-converse}
    \hHm{m}(u^n)
    &\leq \hat{C}_m(\mathcal{N})
    + \phi\!\left(\hat{P}_e^u\right)
    + \frac{5\log s}{m}.
\end{align}
As $\hat{P}_e^u \to 0$ and $m \gg \log s$: $\hHm{m}(u^n) \leq \hat{C}_m(\mathcal{N})$.
\end{theorem}

\begin{proof}
Applying Theorem~\ref{thm:main-tier2} to the pair $(u^n, y^n)$ with the
block-FSM decoder $y^n \mapsto \hat{u}^n$ (block length $m$, state count $s$)
gives:
\[
    \phi\!\left(\hat{P}_e^u\right)
    \geq \hHm{m}(u^n|y^n) - \frac{\log s}{m}.
\]
Since $\hHm{m}(u^n|y^n) = \hHm{m}(u^n) - \hat{I}_m(u^n;y^n)$,
this gives both inequalities in~\eqref{eq:channel-tier1-jscc}.
Now, $\hat{I}_m(u^n;y^n) \leq \hat{I}_m(x^n;y^n) + (4\log s)/m$
by Lemma~\ref{lem:equiv}~\eqref{eq:equiv-MI}, and
$\hat{I}_m(x^n;y^n) \leq \hat{C}_m(\mathcal{N})$
by definition of capacity.
Hence $\hat{I}_m(u^n;y^n) \leq \hat{C}_m(\mathcal{N}) + (4\log s)/m$,
and therefore:
\[
    \phi\!\left(\hat{P}_e^u\right)
    \geq \hHm{m}(u^n) - \hat{C}_m(\mathcal{N}) - \frac{5\log s}{m}.
\]
Rearranging gives~\eqref{eq:jscc-converse}. \qedhere
\end{proof}

\begin{remark}[LZ-level JSCC converse]
\label{rem:jscc-lz}
An analogue of~\eqref{eq:jscc-converse} at the LZ complexity level follows
from the upper sandwiching bound (Theorem~\ref{thm:sandwich}).
Applying~\eqref{eq:sand-upper} to the source sequence $u^n$
with order parameter $m \ll k$:
\[
    \bar{\rho}_k(u^n) \leq \hHm{m}(u^n) + \delta_+(k,m,n,|\mathcal{U}|),
\]
where $\delta_+ \to 0$ as $n \gg k \gg m \gg 1$.
Substituting the converse bound~\eqref{eq:jscc-converse}:
\[
    \bar{\rho}_k(u^n)
    \leq \hat{C}_m(\mathcal{N}) + \phi\!\left(\hat{P}_e^u\right)
    + \frac{5\log s}{m} + \delta_+.
\]
The empirical capacity $\hat{C}_m(\mathcal{N})$ itself satisfies
$\hat{C}_m(\mathcal{N}) \leq C_k(\mathcal{N}) + \delta_+'$
for a second (generally different) redundancy term $\delta_+'$,
obtained by applying the upper sandwich bound to each
$\hat{I}_m(x^n;y^n)$ in the definition of $\hat{C}_m$.
Combining:
\begin{equation}
\label{eq:jscc-converse-lz}
    \bar{\rho}_k(u^n)
    \leq C_k(\mathcal{N}) + \phi\!\left(\hat{P}_e^u\right)
    + \frac{5\log s}{m} + \delta_+ + \delta_+',
\end{equation}
where both redundancy terms vanish in the regime $n \gg k \gg m \gg \log s$.
This is the individual-sequence LZ-level source-channel coding converse:
reliable end-to-end transmission is impossible when
$\bar{\rho}_k(u^n) > C_k(\mathcal{N})$.
\eoe
\end{remark}

\medskip\noindent\textbf{Source-channel separation achievability.}
Theorem~\ref{thm:channel-tier1-jscc} (converse) and
Theorem~\ref{thm:achievability} (LZ achievability, established above)
together yield an individual-sequence analogue of the source-channel
separation theorem.
The achievability direction by separation is as follows.
Suppose $\bar{\rho}_k(u^n) < C_k(\mathcal{N}) - \varepsilon$
for some $\varepsilon > 0$.
First, compress $u^n$ losslessly using LZ78~\cite{Ziv78} to obtain a
message index $w \in \{1,\ldots,M\}$ with
$\log M \leq n\bar{\rho}_k(u^n) + O(\log n) < n(C_k(\mathcal{N}) - \varepsilon/2)$
for large enough $n$.
Then transmit $w$ using the empirical-entropy random code of
Theorem~\ref{thm:achieve-emp} at rate $R = (\log M)/n < C_k(\mathcal{N}) - \varepsilon/2$,
which yields vanishing per-message error frequency for all $v^n \in \mathcal{N}$.
Together with the LZ-level converse~\eqref{eq:jscc-converse-lz},
this constitutes an individual-sequence source-channel separation theorem:
reliable communication is possible if and only if
the LZ complexity of the source does not exceed the LZ channel capacity.

\medskip\noindent\textbf{Why separation holds, and when it is optimal.}
It is worth pausing to interpret the separation theorem correctly.
In classical information theory, the separation principle is in a sense
\emph{bad news}: separate source and channel coding can \emph{always} be
carried out, but the separation theorem merely says that one cannot do
\emph{better} than separation --- only in situations where the principle
fails (e.g., when the source and channel noise are statistically dependent)
can joint source-channel coding outperform separate coding.
The individual-sequence setting makes this even more transparent: here the
noise is adversarial, so the system must be designed against the worst-case
$v^n \in \mathcal{N}$.
Under worst-case noise, the decoupling between source and channel is
complete --- no joint design can exploit correlations between the source and
the noise, since no such correlations are assumed.
Consequently, one cannot expect performance better than that offered by
separate source and channel coding, and the separation theorem is tight.

More precisely, $C_k(\mathcal{N})$ is a worst-case (minimax) quantity
optimized against the hardest noise sequence in $\mathcal{N}$, and
LZ78 produces a nearly uniform bit-string $w$ from $u^n$, decoupling
the source statistics from the channel; the channel code needs only to
transmit $w$ reliably, with no knowledge of the source structure.

For a fixed non-worst-case $v^n$, joint source-channel coding could in
principle exploit the gap between the per-realization capacity and
$C_k(\mathcal{N})$ --- but only with knowledge of which $v^n$ occurred.
Since neither encoder nor decoder knows $v^n$, this improvement is
unavailable in the adversarial setting, and separation incurs no cost.

Note that the absence of a Markov structure $U^n - X^n - Y^n$
in the individual-sequence setting does not by itself
contradict separation.
As an extreme example, consider a modulo-additive channel
with $v^n = u^n$: transmitting $x^n = 0^n$ (rate zero) yields
$y^n = u^n$, so the source is reconstructed perfectly despite
$H = \log\alpha$ and $C = 0$.
There is no contradiction because $v^n = u^n$ is presumably
not the worst noise sequence in $\mathcal{N}$
(and may not even belong to $\mathcal{N}$);
the separation theorem is a statement about the
worst-case $v^n$, not about favourable individual instances.

\section{Discussion}
\label{sec:discussion}

\subsection{The Double Limit and Asymptotic Capacity}

The capacity $C_k(\mathcal{N}_n)$ depends on both $n$ and $k$.
All results in this paper hold at fixed $n$ and $k$; here we
discuss how to pass to the double limit in two distinct settings:
pure channel coding and joint source-channel coding.
The two settings require different limit operations, for
reasons explained below, but both yield the same threshold
$C_\LZ(\calN)$ — which is consistent, since both are answers
to the same underlying question: what is the maximum achievable rate?

\medskip\noindent
\textbf{Pure channel coding.}
In pure channel coding the rate $R$ is a fixed number (not a
sequence in $n$), and the converse
(Theorem~\ref{thm:channel-tier2}) gives, for every $n$ and $k$:
\[
    R \leq C_k(\mathcal{N}_n) + \phi(\hat{P}_e) + \Gamma(n,k),
\]
where $\Gamma(n,k) \to 0$ as $n\to\infty$ for fixed $k$.
Since $R$ does not depend on $n$, taking $\liminf_{n\to\infty}$
of the right-hand side is the natural (and most conservative)
operation:
\[
    R \leq \liminf_{n\to\infty} C_k(\mathcal{N}_n) + \phi(\hat{P}_e).
\]
There is no mismatch between subsequences here: $R$ is fixed,
and the bound holds at every $n$, so it holds in the
$\liminf$.
Taking $\liminf_{k\to\infty}$ then gives:
\[
    R \leq \liminf_{k\to\infty}\liminf_{n\to\infty}
    C_k(\mathcal{N}_n) + \phi(\hat{P}_e).
\]
This motivates the pessimistic definition of asymptotic capacity:
\begin{equation}
\label{eq:asym-cap}
    C_\LZ(\mathcal{N}) \;\dfn\;
    \liminf_{k\to\infty}\liminf_{n\to\infty} C_k(\mathcal{N}_n).
\end{equation}
The asymptotic converse is then simply $R \leq C_\LZ(\calN) + \phi(\hat{P}_e)$.

For the achievability: if $R < C_\LZ(\calN)$, then for all
sufficiently large $k$ and $n$, $R < C_k(\calN_n) - \eps/2$
for some $\eps > 0$, and Theorem~\ref{thm:achievability}
gives $\mathbb{E}_\omega[\hat{P}_e] \to 0$.
Together: reliable communication at rate $R$ is possible if
and only if $R \leq C_\LZ(\calN)$
(with the boundary case $R = C_\LZ(\calN)$ left open, as is
standard in the LZ framework).

A necessary consistency check: the paper has two capacity
expressions at finite $n$ and $k$ --- the empirical-entropy
capacity $\hat{C}_m(\calN_n)$ (Definition~\ref{def:capacity})
and the LZ capacity $C_k(\calN_n)$
(Definition~\ref{def:capacity-k}) --- and they must agree in
the double limit:
\[
    \liminf_{k\to\infty}\liminf_{n\to\infty} \hat{C}_m(\calN_n)
    \;=\;
    \liminf_{k\to\infty}\liminf_{n\to\infty} C_k(\calN_n)
    \;=\; C_\LZ(\calN).
\]
This is not a coincidence: both are operational thresholds for
the same channel coding problem, so they \emph{must} agree
asymptotically. The sandwich relation between
$\hat{C}_m(\calN_n)$ and $C_k(\calN_n)$ (via the upper and
lower sandwiching bounds of Section~\ref{sec:lz-capacity}) is
what makes this precise at finite $n$ and $k$, and guarantees
agreement in the limit.

\medskip\noindent
\textbf{Joint source-channel coding.}
In JSCC, the source LZ complexity $\bar{\rho}_k(x^n)$ enters
the picture alongside the channel capacity, and the two must
be compared \emph{at the same $n$}.
Here the relevant object is the difference:
\begin{equation}
    D_k(n) \dfn \bar{\rho}_k(x^n) - C_k(\mathcal{N}_n).
\end{equation}
A natural but incorrect approach would be to compare
$\limsup_n \bar{\rho}_k(x^n)$ with $\liminf_n C_k(\calN_n)$
separately, paralleling the ZL78 definition of compressibility
$\bar{\rho}(x) \dfn \limsup_{n\to\infty}\rho_\LZ(x^n)$.
This is wrong because the two limits may be attained on
\emph{different} subsequences: at some $n_0$, $\bar{\rho}_k(x^{n_0})$
may be large while $C_k(\calN_{n_0})$ is also large (a
good channel length cancels a bad source length, giving a
vacuous bound); at another $n_1$, $\bar{\rho}_k(x^{n_1})$ may
be small while $C_k(\calN_{n_1})$ is also small (a bad
channel length cancels a good source length).
Taking separate limits misses this coupling entirely.

The correct formulation limsups the \emph{difference} $D_k(n)$
at the same $n$:
\begin{equation}
\label{eq:asymp-converse}
    \limsup_{n\to\infty} D_k(n) \leq \phi(\hat{P}_e),
\end{equation}
obtained by taking $\limsup_n$ of the LZ-level JSCC converse
\eqref{eq:jscc-converse-lz} and noting that redundancy terms
vanish.
The achievability condition is $\limsup_n D_k(n) < 0$.

Taking $\limsup_{k\to\infty}$ (since whether
$\lim_k \limsup_n D_k(n)$ exists is open), and denoting:
\begin{equation}
\label{eq:D-limit}
    D \;\dfn\; \limsup_{k\to\infty}\limsup_{n\to\infty} D_k(n),
\end{equation}
the asymptotic JSCC coding theorem is:
$D \leq \phi(\hat{P}_e)$ (converse) and
$D < 0 \Rightarrow \mathbb{E}_\omega[\hat{P}_e]\to 0$ (achievability).
For reliable communication ($\hat{P}_e \to 0$) this collapses to
$D \leq 0$, i.e.,
\[
    \limsup_{k\to\infty}\limsup_{n\to\infty}
    \bigl[\bar{\rho}_k(x^n) - C_k(\mathcal{N}_n)\bigr] \leq 0,
\]
which is the individual-sequence analogue of the source-channel
separation condition $\bar{\rho}(x) \leq C_\LZ(\calN)$.

\medskip\noindent
\textbf{Relation to finite-$n$ results.}
All theorems in this paper hold for fixed $n$ and $k$,
giving per-sequence bounds without any limiting argument.
The asymptotic formulations above are the natural summaries
that emerge from the double limit.

\subsection{Bridging the Gap Between the Converse and the Achievability}
\label{sec:gap}

There is an apparent asymmetry between the converse and the
achievability on the decoder side.
The converse (Theorem~\ref{thm:channel-tier2}) is stated for a
block-FSM decoder with $s$ states and block length $m$, which processes
$y^n$ sequentially in blocks of length $m$ with bounded inter-block memory.
The achievability decoder, by contrast, compares metrics computed over the
\emph{entire} sequence $y^n$: it selects the codeword minimizing
$\bar{\rho}_k(\tilde{x}^n \mid y^n)$ globally, and in that sense requires
observing all $n$ output symbols before producing any decision.
This global look at $y^n$ is not a block-FSM operation at the $n$-symbol
scale, regardless of the fact that the metric $\bar{\rho}_k$ is computed
in $k$-blocks internally.

The gap between the two decoder classes is closed by passing to
\emph{super-blocks} of $T$ independent transmissions of length $n$.
Over a super-block of length $nT$, the achievability decoder operates
independently on each of the $T$ sub-blocks of length $n$ (since the
codebook and channel are both applied independently per block), so the
global $nT$-symbol decoder decomposes into $T$ independent $n$-symbol
decisions.
In the limit $T \to \infty$ (with $n$ fixed), this is equivalent to
an $n$-block FSM with $s = 1$ state (no inter-block memory), which falls
within the converse's block-FSM class with block length $m = n$ and $s = 1$.
The correction term then satisfies $(\log s)/m = 0$ identically,
and $\Gamma_0 = \delta_+(k,m,n,\alpha\beta) \to 0$
as $n \gg k \gg m \gg 1$ (with $m = n$ being the ``super-block'' block length
passed to the converse).

On the encoder side, the converse imposes no finite-state
constraint; the achievability's lookup-table block encoder
(which maps each message $i$ to a full $n$-length codeword
$X(i)$ drawn uniformly from the $k$-type class $\calT_k(P^*_{X^k})$)
is therefore compatible with the converse without any
additional restriction.

In the double limit $n \to \infty$ followed by $k \to \infty$,
the converse bound is tight: $\Gamma_0 \to 0$ and
$C_k(\mathcal{N}_n) \to C_\LZ(\mathcal{N})$,
the individual-sequence LZ capacity 
defined in~\eqref{eq:asym-cap} as
$$\liminf_{k\to\infty}\liminf_{n\to\infty} C_k(\mathcal{N}_n).$$
Whether the converse extends to arbitrary (unsynchronized) decoders that need not operate block-by-block is also an open problem.

\subsection{Does Feedback Increase the Individual-Sequence LZ Capacity?}

An important open problem is whether feedback from the
decoder to the encoder can increase the individual-sequence
LZ capacity $C_k(\mathcal{N})$.
In the classical probabilistic setting, feedback does not increase capacity for memoryless
channels, but it can for channels with memory~\cite{CT91}.
In the individual-sequence setting, the noise sequence
$v^n$ introduces temporal correlations across channel
uses, so feedback may strictly increase the
individual-sequence LZ capacity.
If so, the achievability schemes of Shayevitz, Feder,
and Lomnitz~\cite{SF11,LF14,LF13b}, which use feedback,
may be achieving a genuinely larger capacity than
our converse (Theorem~\ref{thm:channel-tier2}) bounds.
Characterizing the individual-sequence LZ capacity with
feedback is an open problem.

\end{document}